\documentclass[10pt, a4paper]{article}
\pdfoutput=1
\usepackage{jheppub}
\usepackage[english]{babel}
\usepackage[latin1]{inputenc}
\usepackage{amsmath}
\usepackage{amssymb}
\usepackage{bbm}
\usepackage{graphicx}
\usepackage{xcolor}
\begin{document}

\title{Quantum symmetries in 2+1 dimensions: Carroll, (a)dS-Carroll, Galilei and (a)dS-Galilei}

\author{Tomasz Trze\'{s}niewski}
\affiliation{Institute of Theoretical Physics, University of Wroc\l{}aw, pl.\ M.\ Borna 9, 50-204 Wroc\l{}aw, Poland}
\emailAdd{tomasz.trzesniewski@uwr.edu.pl}

\abstract{There is a surge of research devoted to the formalism and physical manifestations of non-Lorentzian kinematical symmetries, which focuses especially on the ones associated with the Galilei and Carroll relativistic limits (the speed of light taken to infinity or to zero, respectively). The investigations have also been extended to quantum deformations of the Carrollian and Galilean symmetries, in the sense of (quantum) Hopf algebras. The case of 2+1 dimensions is particularly worth to study due to both the mathematical nature of the corresponding (classical) theory of gravity, and the recently finalized classification of all quantum-deformed algebras of spacetime isometries. Consequently, the list of all quantum deformations of (anti-)de Sitter-Carroll algebra is immediately provided by its well-known isomorphism with either Poincar\'{e} or Euclidean algebra. Quantum contractions from the (anti-)de Sitter to (anti-)de Sitter-Carroll classification allow to almost completely recover the latter. One may therefore conjecture that the analogous contractions from the (anti-)de Sitter to (anti-)de Sitter-Galilei $r$-matrices provide (almost) all coboundary deformations of (anti-)de Sitter-Galilei algebra. This scheme is complemented by deriving (Carrollian and Galilean) quantum contractions of deformations of Poincar\'{e} algebra, leading to coboundary deformations of Carroll and Galilei algebras.}

\maketitle

\section{Introduction}

Similarly as symmetries of Minkowski spacetime are specified by Poincar\'{e} algebra (i.e., this spacetime is the homogeneous space of Poincar\'{e} group), the ``space-time'' of classical non-relativistic mechanics can be defined as the homogeneous space of the group generated by Galilei algebra. Furthermore, both algebras belong to the much richer realm of kinematical algebras -- a particular type of Lie algebras -- which became the object of physical interest due to the seminal work of Bacry and L\'{e}vy-Leblond \cite{Bacry:1968ps}. The classification of such algebras in 3+1 spacetime dimensions has subsequently been completed in \cite{Bacry:1986cy}, but only recently in the cases of $>$3+1 \cite{Figueroa:2018hy} and 2+1 dimensions \cite{Andrzejewski:2018ks}, using the method of deformations (which should not be confused with quantum deformations discussed below). Kinematics turns out to be even more diverse at the level of the homogeneous spaces of kinematical groups \cite{Figueroa:2019ss}. Nowadays, this subject matter is experiencing increased attention, as demonstrated by many papers, including last year's reviews \cite{Bergshoeff:2023ar,Figueroa:2022ns} and references therein. In the same vein, research has been directed to the non-Lorentzian versions of BMS (Bondi-van der Burg-Metzner-Sachs) \cite{Perez:2021ay,Fuentealba:2022as} and other algebras describing asymptotic symmetries of spacetime, which are also interrelated by appropriate deformations, see e.g. \cite{Parsa:2019os,Safari:2019os}. 

A particular interest is attracted by Carroll algebra and its generalizations, corresponding to the ultra-relativistic limit (the speed of light taken to zero) of Poincar\'{e}, (anti-)de Sitter, and other algebras. Particles \cite{Bergshoeff:2014ds} or strings \cite{Cardona:2016ds} on a Carrollian manifold (the homogeneous space of Carroll group) have trivial dynamics unless they are interacting or coupled to a gauge field, cf. \cite{Marsot:2022pn,Marsot:2023hs}. As one should expect, the Carrollian algebraic and geometric structures are to a certain extent ``dual'' \cite{Duval:2014ce,Figueroa:2023ly} to their Galilean counterparts, which are obtained in the non-relativistic limit (the speed of light taken to infinity; strictly speaking, it is the limit leading to Galilean relativity). This extends \cite{Bergshoeff:2017cy} to the Carroll and Galilei limits of the theory of gravity but the relation is not actually reciprocal, e.g. there exist two inequivalent Carrollian contractions of general relativity, the so-called: electric \cite{Henneaux:1979gs} and magnetic \cite{Henneaux:2021cs}. An expansion around the Carroll limit \cite{Hansen:2022cy} is related to the strong-coupling expansion of general relativity \cite{Isham:1976sy}, as well as the BKL conjecture for spacetime singularities \cite{Belinsky:1970oy}. An interesting parallel in this context is that the Carroll limit makes light cones collapse into temporal lines (hence its alternative name: the ultra-local limit), while the evolution towards a singularity is associated with asymptotic silence, i.e. the shrinking of particle horizons to zero \cite{Andersson:2005as} (see also \cite{Mielczarek:2013ay,Mielczarek:2017se}). On the other hand \cite{Duval:2014ce,Figueroa:2019ss}, Carroll/(anti-)de Sitter-Carroll algebra generates symmetries of a null hypersurface in the Lorentzian spacetime one dimension higher, i.e. such surfaces are Carrollian manifolds. This concerns in particular (Dodgson) gravitational waves \cite{Morand:2020es}, which can be foliated by the latter, while Carroll group (with broken rotations) describes \cite{Duval:2017cs} their isometries. Another example is the event horizon of a black hole, at least if one considers it within the so-called membrane paradigm \cite{Donnay:2019cn}. Moreover, BMS group is isomorphic to a conformal extension of Carroll group of one dimension less \cite{Duval:2014cs,Ciambelli:2019cs}. 

The assortment of symmetries available for study in physical theories can also be enlarged by their (quantum) deformations, described in the formalism of Hopf algebras, which provides a generalization of the notion of an associative algebra. (Quantum) homogeneous spaces of such quantum-deformed algebras of symmetries turn out to have noncommutative geometry. The most extensively studied example is $\kappa$-Poincar\'{e} (Hopf) algebra \cite{Lukierski:1991qa,Lukierski:1992ny} (which can be defined in any number of dimensions \cite{Maslanka:1993tp}), associated with $\kappa$-Minkowski (noncommutative) spacetime \cite{Majid:1994by}. Their counterparts for the case of non-zero cosmological constant are given by $\kappa$-(anti-)de Sitter algebra and spacetime \cite{Ballesteros:2017ts,Ballesteros:2019tf} (first introduced in 2+1 dimensions \cite{Ballesteros:1994qh,Ballesteros:2017ns}). One of the most recent advances is the construction of deformations of BMS algebras \cite{Borowiec:2019ky,Borowiec:2021bs,Borowiec:2021ds}. 

In the case of kinematical symmetries, the Hopf algebra is a deformation of (the universal enveloping algebra of) a Lie algebra, while its infinitesimal version is a Lie bialgebra, generating a Poisson-Lie group. A particular kind of the latter is the coboundary Lie bialgebra, characterized by a distinguished element of the tensor product of its two copies known as a classical $r$-matrix (see Sec.~\ref{sec:2.0}). If a given Lie algebra is semisimple or inhomogeneous (pseudo-)orthogonal, all possible bialgebras are coboundary and hence it suffices to classify the corresponding $r$-matrices in order to find all quantum (i.e. Hopf-algebraic) deformations. So far, such a classification has only been completed in 2+1-dimensions -- for Poincar\'{e} and (inhomogeneous) Euclidean algebras \cite{Stachura:1998ps}, as well as (Lorentzian) (anti-)de Sitter and Euclidean (anti-)de Sitter algebras \cite{Zakrzewski:1994pp,Borowiec:2016qg,Borowiec:2017ag,Borowiec:2017bs}. The latter has been achieved by treating $\mathfrak{so}(4)$, $\mathfrak{so}(3,1)$ and $\mathfrak{so}(2,2)$ (as well as $\mathfrak{o}(2;\mathbbm{H})$) as real forms of the algebra $\mathfrak{so}(4;\mathbbm{C})$. In our previous work \cite{Kowalski:2020qs}, we investigated how $r$-matrices of all these kinematical algebras can be related by the quantum \.{I}n\"{o}n\"{u}-Wigner contractions, i.e. by the appropriately taken limits of the cosmological constant approaching zero. Meanwhile, the classification for Poincar\'{e} algebra in 3+1 dimensions \cite{Zakrzewski:1997pp} is largely complete but there are still some missing cases, while even less is known for (anti-)de Sitter algebra \cite{Ballesteros:2017ts,Ballesteros:2022ns}. 

What is the motivation for considering such structures? Some form of deformed relativistic symmetries and noncommutativity of the spacetime geometry is widely argued to emerge in description of the quantum regime of gravity \cite{Oriti:2009ay}, especially within quantum gravity phenomenology (see a last year's review \cite{Addazi:2022qw}), where it is an alternative to a more far-reaching idea of symmetry breaking. Other area of applications has been found in string theory and the AdS/CFT correspondence -- the Yang-Baxter deformations of strings or, more generally, sigma models, see e.g. \cite{Pachol:2016qg,Osten:2017as,Idiab:2022yg} or a review \cite{Hoare:2022is}. Last but not least, (infinitesimally) deformed algebras are present already at the classical level in the theory of (2+1)-dimensional gravity. The latter can be recast \cite{Achucarro:1986as,Witten:1988dm} as Chern-Simons theory on spacetime that is locally Minkowski or (anti-)de Sitter, depending on the cosmological constant, with the corresponding group of local isometries as the gauge group. In this framework, it has been shown that the Poisson structure of phase space (the space of flat Cartan connections modulo gauge transformations) is determined by a classical $r$-matrix compatible with a given Chern-Simons action \cite{Fock:1999px}, which also turns the local isometry group into a Poisson-Lie group and provides a direct link with the Hopf-algebraic quantum deformations of symmetries \cite{Meusburger:2008qr,Meusburger:2009gy,Schroers:2011qh}. The relevant $r$-matrices for each gauge group have already been classified \cite{Osei:2018cy,Kowalski:2020qs} (see also \cite{Osei:2012oy,Rosati:2017kt,Ballesteros:2013dy,Ballesteros:2019te}) but it remains unclear whether they lead to physically distinct theories. 

As we discussed at the beginning, the non-Lorentzian kinematical algebras (i.e., other ones than Poincar\'{e} or (anti-)de Sitter) are not a mere curiosity but play various roles in the theory of gravity. In the case of 2+1 dimensions, the Chern-Simons has also been recently generalized to spacetimes modelled by the homogeneous space of an arbitrary kinematical group, which then becomes the gauge group \cite{Matulich:2019ln} (see also \cite{Kowalski:2014dy,Trzesniewski:2018ey,Trzesniewski:2023gs} for a different way in which deformed Carrollian dynamics can apparently arise in (2+1)d gravity). Theories with different gauge groups are related by the proper limits, e.g. the speed of light going to zero or to infinity. Therefore, the natural question to ask is what are the (quantum) deformations of non-Lorentzian kinematical algebras, as well as what happens with deformations of the Lorentzian algebras in the non-Lorentzian limits. Some examples of such deformations were studied in e.g. \cite{Giller:1992ma,Daszkiewicz:2008cs,Daszkiewicz:2019cs,Ballesteros:2020ts}, while a more systematic approach was given in \cite{Ballesteros:1994qh,Ballesteros:1994fs,Gutierrez:2021cs}. In particular, \cite{Ballesteros:2020ts} (as well as \cite{Ballesteros:2021is}) contains a derivation of the (quantum) Carrollian and Galilean contraction limits of Lie bialgebras and Hopf algebras for the above-mentioned (timelike) $\kappa$-deformations of Poincar\'{e} and (anti-)de Sitter algebras in 3+1 dimensions. 


The plan of this paper is as follows. Sec.~\ref{sec:2.0} recalls the concept of classical $r$-matrices and then their complete classification for Poincar\'{e} and (anti-)de Sitter algebras in 2+1 dimensions, characterizing all possible Hopf-algebraic deformations of these Lorentzian algebras. Sec.~\ref{sec:3.0} discusses the definitions and relevant properties of the Carrollian and Galilean versions of (undeformed) Poincar\'{e} and (anti-)de Sitter algebras (the paper also contains Appendix concerning the relation between Carroll algebra and Poincar\'{e} algebra one dimension higher), followed by a derivation of the Carroll and Galilei limits of deformations of Poincar\'{e} algebra via the procedure of quantum contractions. In Subsec.~\ref{sec:4.1}, we completely classify deformations of (anti-)de Sitter-Carroll algebra, with the help of isomorphisms between anti-de Sitter-Carroll and Poincar\'{e} algebra, and between de Sitter-Carroll and Euclidean algebra. Subsequently, in the rest of Sec.~\ref{sec:4.0}, we show that all such deformations (up to a few terms missing in some classes of $r$-matrices) can be recovered as quantum contraction limits of deformed (anti-)de Sitter algebras. Sec.~\ref{sec:5.0} is similarly devoted to these deformations of (anti-)de Sitter-Galilei algebra that can be derived via quantum contractions from the (anti-)de Sitter case (we conjecture that they provide the major part of the unknown complete classification). Sec.~\ref{sec:6.0} ends it with a summary of our results and a discussion of their possible applications, as well as some comments on the especially interesting cases of deformations. In particular, we observe that deformations of the Carrollian and Galilean algebras are in a sense milder than deformations of Lorentzian algebras, in the manner that reflects the crucial features of Carrollian and Galilean kinematics (the ultralocality and absolute time, respectively).

\section{Classical $r$-matrices (in 2+1 dimensions)}\label{sec:2.0}

Let us start with a very brief reminder of the necessary mathematical preliminaries. We refer the Reader to the direct predecessors of this paper \cite{Borowiec:2017bs,Kowalski:2020qs} for more details, or to e.g. \cite{Chari:1994as,Majid:1995fy} for an in-depth discussion of the subject. 

A Lie algebra $\mathfrak{g}$ with a Lie bracket $[\cdot,\cdot]: \mathfrak{g} \otimes \mathfrak{g} \mapsto \mathfrak{g}$ becomes a Lie bialgebra when equipped with a compatible structure of a (Lie) cobracket $\delta(\cdot): \mathfrak{g} \mapsto \mathfrak{g} \otimes \mathfrak{g}$. A particularly interesting type of Lie bialgebras are the coboundary ones, for which the cobracket is determined by an element $r \in \mathfrak{g} \wedge \mathfrak{g}$ called a classical (antisymmetric) $r$-matrix:
\begin{align}\label{eq:10.00}
\forall g \in \mathfrak{g}:\ \delta(g) = {\rm ad}_g r = [g, r^{(1)}] \otimes r^{(2)} + r^{(1)} \otimes [g, r^{(2)}]\,.
\end{align}
(In Sweedler notation, $r = \sum_i r^{(1)}_i \otimes r^{(2)}_i \equiv r^{(1)} \otimes r^{(2)}$.) It follows that a given $r$-matrix is defined up to the terms that do not change the corresponding cobracket, i.e. up to the elements of $\mathfrak{g} \wedge \mathfrak{g}$ that commute with every $g \in \mathfrak{g}$. We may call such $\mathfrak{g}$-invariant tensors the antisymmetric split-Casimirs, since the name of split Casimirs is used in the more general case of the $\mathfrak{g}$-invariant elements of $\mathfrak{g} \otimes \mathfrak{g}$. Furthermore, if two $r$-matrices can be transformed into each other by (mutually inverse) automorphisms of $\mathfrak{g}$, they naturally determine isomorphic Lie bialgebras and can be identified as belonging to the same equivalence class. In order to stress this, we will sometimes use the name ``$r$-matrix class'' instead of just ``$r$-matrix''. 

The equivalent definition of a classical $r$-matrix for a Lie algebra $\mathfrak{g}$ is that it solves the (classical) Yang-Baxter equation
\begin{align}\label{eq:10.01}
[[r,r]] = t\, \Omega\,, \quad t \in \mathbbm{C}\,.
\end{align}
$[[\cdot,\cdot]]$ denotes the Schouten bracket, defined as
\begin{align}\label{eq:10.02}
[[a \wedge b,c \wedge d]] := a \wedge \left([b,c] \wedge d + c \wedge [b,d]\right) - b \wedge \left([a,c] \wedge d + c \wedge [a,d]\right),
\end{align}
while $\Omega$ is a $\mathfrak{g}$-invariant element of $\mathfrak{g} \otimes \mathfrak{g} \otimes \mathfrak{g}$. If the Yang-Baxter equation is homogeneous, i.e. $t = 0$, its solutions are called the triangular $r$-matrices and correspond to twist quantizations of the bialgebra $\mathfrak{g}$; if the Yang-Baxter equation is inhomogeneous (also known as the modified YB equation), i.e. $t \neq 0$, its solutions are called the quasitriangular $r$-matrices and correspond to quasitriangular Hopf algebras describing quantizations of $\mathfrak{g}$. 

\subsection{The classification for Poincar\'{e} algebra}

The 2+1-dimensional Poincar\'{e} algebra $\mathfrak{iso}(2,1) = \mathfrak{so}(2,1) \vartriangleright\!\!< \mathbbm{R}^{2,1}$ in the orthogonal basis, which makes explicit its semidirect product structure, has the brackets:
\begin{align}\label{eq:32.03}
[{\cal J}_\mu,{\cal J}_\nu] = \epsilon_{\mu\nu}^{\ \ \sigma} {\cal J}_\sigma\,, \qquad [{\cal J}_\mu,{\cal P}_\nu] = \epsilon_{\mu\nu}^{\ \ \sigma} {\cal P}_\sigma\,, \qquad [{\cal P}_\mu,{\cal P}_\nu] = 0
\end{align}
($\mu = 0,1,2$, we set $\epsilon_{012} = 1$ and rise indices with Minkowski metric $(1,-1,-1)$). The classification of classical $r$-matrices for this algebra, completed by Stachura in \cite{Stachura:1998ps}, can be written down as the following list of disjoint multi-parameter families (up to automorphisms of the algebra; as presented in \cite{Stachura:1998ps}, the dependence on most of the parameters can be simplified or even eliminated by acting with dilations, but the parameters are relevant from the physical point of view):
\begin{align}\label{eq:60.01}
r_1(\chi,\gamma) &= \chi\, ({\cal J}_0 + {\cal J}_1) \wedge {\cal J}_2 + \gamma\, ({\cal J}_0 \wedge {\cal P}_0 - {\cal J}_1 \wedge {\cal P}_1 - {\cal J}_2 \wedge {\cal P}_2)\,, \nonumber\\
r_2(\gamma,\eta;\theta_{01},\theta_{12},\theta_{20}) &= \gamma\, ({\cal J}_0 \wedge {\cal P}_2 - {\cal J}_2 \wedge {\cal P}_0) + \eta\, {\cal J}_1 \wedge {\cal P}_1 + r_8(\theta_{01},\theta_{12},\theta_{20})\,, \nonumber\\
r_3(\gamma,\eta;\theta_{01},\theta_{12},\theta_{20}) &= \gamma\, ({\cal J}_1 \wedge {\cal P}_2 - {\cal J}_2 \wedge {\cal P}_1) + \eta\, {\cal J}_0 \wedge {\cal P}_0 + r_8(\theta_{01},\theta_{12},\theta_{20})\,, \nonumber\\
r_4(\chi,\varsigma;\theta_{01},\theta_{12},\theta_{20}) &= \chi \left({\cal J}_1 \wedge ({\cal P}_0 + {\cal P}_2) - ({\cal J}_0 + {\cal J}_2) \wedge {\cal P}_1\right) + \varsigma\, ({\cal J}_0 + {\cal J}_2) \wedge ({\cal P}_0 + {\cal P}_2) \nonumber\\
&+ r_8(\theta_{01},\theta_{12},\theta_{20})\,, \nonumber\\
r_5(\chi;\theta_{01},\theta_{12},\theta_{20}) &= \chi\, {\cal J}_1 \wedge ({\cal P}_0 + {\cal P}_2) + r_8(\theta_{01},\theta_{12},\theta_{20})\,, \nonumber\\
r_6(\gamma,\varsigma;\theta_{01},\theta_{12},\theta_{20}) &= \gamma\, ({\cal J}_0 \wedge {\cal P}_2 - {\cal J}_2 \wedge {\cal P}_0 + {\cal J}_1 \wedge {\cal P}_1) + \varsigma\, ({\cal J}_0 + {\cal J}_2) \wedge ({\cal P}_0 + {\cal P}_2) \nonumber\\
&+ r_8(\theta_{01},\theta_{12},\theta_{20})\,, \nonumber\\
r_7(\gamma) &= \gamma\, ({\cal J}_0 \wedge {\cal P}_0 - {\cal J}_1 \wedge {\cal P}_1 - {\cal J}_2 \wedge {\cal P}_2)\,, \nonumber\\
r_8(\theta_{01},\theta_{12},\theta_{20}) &= \theta_{01} {\cal P}_0 \wedge {\cal P}_1 + \theta_{12} {\cal P}_1 \wedge {\cal P}_2 + \theta_{20} {\cal P}_2 \wedge {\cal P}_0\,.
\end{align}
The deformation parameters are assumed to be restricted by a few conditions: $\gamma \neq 0 \vee \eta \neq 0$ and $\chi \neq 0 \vee \varsigma \neq 0$, as well as $\gamma \neq \eta$ (in $r_2$), $\chi \neq 0$ (in $r_1$ and $r_5$) and $\gamma \neq 0$ (in $r_6$ and  $r_7$). Otherwise, different (classes of) $r$-matrices would overlap for certain values of the parameters, e.g. $r_2(\gamma,\eta = \gamma) = r_6(\gamma,\varsigma = 0)$. 

Actually, $\mathfrak{iso}(2,1)$ automorphisms allow us \cite{Stachura:1998ps} to get rid of some terms depending only on ${\cal P}_\mu$ generators (i.e. the terms that are themselves $r$-matrices of class $r_8$), and bring the classification (\ref{eq:60.01}) to an even simpler form:
\begin{align}\label{eq:60.01b}
r_1(\chi,\gamma) &= \chi\, ({\cal J}_0 + {\cal J}_1) \wedge {\cal J}_2 + \gamma\, ({\cal J}_0 \wedge {\cal P}_0 - {\cal J}_1 \wedge {\cal P}_1 - {\cal J}_2 \wedge {\cal P}_2)\,, \nonumber\\
r_2(\gamma,\eta;\theta_{20}) &= \gamma\, ({\cal J}_0 \wedge {\cal P}_2 - {\cal J}_2 \wedge {\cal P}_0) + \eta\, {\cal J}_1 \wedge {\cal P}_1 + \theta_{20} {\cal P}_2 \wedge {\cal P}_0\,, \nonumber\\
r_3(\gamma,\eta;\theta_{12}) &= \gamma\, ({\cal J}_1 \wedge {\cal P}_2 - {\cal J}_2 \wedge {\cal P}_1) + \eta\, {\cal J}_0 \wedge {\cal P}_0 + \theta_{12} {\cal P}_1 \wedge {\cal P}_2\,, \nonumber\\
r_4(\chi,\varsigma;\theta) &= \chi \left({\cal J}_1 \wedge ({\cal P}_0 + {\cal P}_2) - ({\cal J}_0 + {\cal J}_2) \wedge {\cal P}_1\right) + \varsigma\, ({\cal J}_0 + {\cal J}_2) \wedge ({\cal P}_0 + {\cal P}_2) \nonumber\\
&+ \theta\, ({\cal P}_0 - {\cal P}_2) \wedge {\cal P}_1\,, \nonumber\\
r_5(\chi;\theta_{01},\theta_{12}) &= \chi\, {\cal J}_1 \wedge ({\cal P}_0 + {\cal P}_2) + \theta_{01} {\cal P}_0 \wedge {\cal P}_1 + \theta_{12} {\cal P}_1 \wedge {\cal P}_2\,, \nonumber\\
r_6(\gamma,\varsigma;\theta,\theta_{20}) &= \gamma\, ({\cal J}_0 \wedge {\cal P}_2 - {\cal J}_2 \wedge {\cal P}_0 + {\cal J}_1 \wedge {\cal P}_1) + \varsigma\, ({\cal J}_0 + {\cal J}_2) \wedge ({\cal P}_0 + {\cal P}_2) \nonumber\\
&+ \theta\, ({\cal P}_0 - {\cal P}_2) \wedge {\cal P}_1 + \theta_{20} {\cal P}_2 \wedge {\cal P}_0\,, \nonumber\\
r_7(\gamma) &= \gamma\, ({\cal J}_0 \wedge {\cal P}_0 - {\cal J}_1 \wedge {\cal P}_1 - {\cal J}_2 \wedge {\cal P}_2)\,, \nonumber\\
r_8(\theta_{01},\theta_{12},\theta_{20}) &= \theta_{01} {\cal P}_0 \wedge {\cal P}_1 + \theta_{12} {\cal P}_1 \wedge {\cal P}_2 + \theta_{20} {\cal P}_2 \wedge {\cal P}_0\,.
\end{align}
Since $\mathfrak{iso}(2,1)$ is an inhomogeneous pseudo-orthogonal algebra in $\geq$3 dimensions, a classical $r$-matrix corresponds to each of its Hopf-algebraic deformations. In particular, if the $\theta,\theta_{\mu\nu}$ parameters vanish, $r_3$, $r_2$ and $r_4$ determine the (twisted) time-, space- and lightlike $\kappa$-deformations, respectively, while $r_6$ is a special combination of the twisted spacelike $\kappa$-deformation and the same twist as for the lightlike $\kappa$-deformation. Moreover, $r_7$ is associated with a particular Drinfeld double of 2+1-dimensional Lorentz algebra; certain special cases of $r_6$ (in which we included a subclass of $r_2$ with $\gamma = \eta$, cf. (\ref{eq:60.01})) and $r_1$ are associated with seven other possible Drinfeld double structures \cite{Ballesteros:2019te}. 

Every $r$-matrix is also a solution of the Yang-Baxter equation (\ref{eq:10.01}), whose RHS is a (possibly non-zero) invariant of the algebra under consideration. The $\mathfrak{iso}(2,1)$ algebra has two linearly independent invariants, $\epsilon^{\mu\nu\sigma} {\cal J}_\mu \wedge {\cal P}_\nu \wedge {\cal P}_\sigma$ and ${\cal P}_0 \wedge {\cal P}_1 \wedge {\cal P}_2$ (let us note that there is no analogue of the second one for Poincar\'{e} algebra in $\neq 2 + 1$ dimensions). Indeed, calculating the Schouten brackets we verify that the ones given in (\ref{eq:60.01}) (or (\ref{eq:60.01b})) satisfy the following inhomogeneous equations:
\begin{align}\label{eq:60.02}
[[r_1,r_1]] = [[r_7,r_7]] &= -\gamma^2 \epsilon^{\mu\nu\sigma} {\cal J}_\mu \wedge {\cal P}_\nu \wedge {\cal P}_\sigma\,, \nonumber\\
[[r_2,r_2]] = [[r_6,r_6]] &= -\gamma^2 \epsilon^{\mu\nu\sigma} {\cal J}_\mu \wedge {\cal P}_\nu \wedge {\cal P}_\sigma + 4\gamma\, \theta_{20}\, {\cal P}_0 \wedge {\cal P}_1 \wedge {\cal P}_2\,, \nonumber\\
[[r_3,r_3]] &= \gamma^2 \epsilon^{\mu\nu\sigma} {\cal J}_\mu \wedge {\cal P}_\nu \wedge {\cal P}_\sigma + 4\gamma\, \theta_{12}\, {\cal P}_0 \wedge {\cal P}_1 \wedge {\cal P}_2\,, \nonumber\\
\tfrac{1}{2}\, [[r_4,r_4]] = [[r_5,r_5]] &= 2\chi\, (\theta_{01} + \theta_{12})\, {\cal P}_0 \wedge {\cal P}_1 \wedge {\cal P}_2\,,
\end{align}
as well as the homogeneous equation in the case of $r_8$. Moreover, $r_4$ and $r_5$ satisfy the homogeneous Yang-Baxter equation if all of their relevant $\theta_{\mu\nu}$ parameters are zero or $\theta_{01} = -\theta_{12}$, while for $r_2$ and $r_3$ it happens when both $\gamma$ and the relevant $\theta_{\mu\nu}$ parameters are zero, and for $r_1$ -- when $\gamma = 0$ ($\gamma$ in $r_6$ and $r_7$ is restricted to be non-zero, as we mentioned below (\ref{eq:60.01})).

\subsection{The classification for (anti-)de Sitter algebra}\label{sec:2.2}

The algebra (\ref{eq:32.03}) can be straightforwardly generalized to (2+1-dimensional) de Sitter and anti-de Sitter algebras, $\mathfrak{so}(3,1)$ and $\mathfrak{so}(2,2)$, corresponding respectively to the cosmological constant $\Lambda > 0$ and $\Lambda < 0$, whose brackets are written in a unified fashion as:
\begin{align}\label{eq:32.03a}
[{\cal J}_\mu,{\cal J}_\nu] = \epsilon_{\mu\nu}^{\ \ \sigma} {\cal J}_\sigma\,, \qquad [{\cal J}_\mu,{\cal P}_\nu] = \epsilon_{\mu\nu}^{\ \ \sigma} {\cal P}_\sigma\,, \qquad [{\cal P}_\mu,{\cal P}_\nu] = -\Lambda\, \epsilon_{\mu\nu}^{\ \ \sigma} {\cal J}_\sigma\,.
\end{align}
The complete classification of classical $r$-matrices for both algebras (up to their automorphisms) has been provided in \cite{Borowiec:2017bs}, building upon the older work of \cite{Zakrzewski:1994pp}. We will present it in the form from our previous paper \cite{Kowalski:2020qs}, where it has been expressed in the basis (\ref{eq:32.03a}), but changing the metric signature to the one used in this paper and redefining some of the parameters. The parameters in \cite{Kowalski:2020qs} were actually inherited from the classification of $r$-matrices for the $\mathfrak{o}(4;\mathbbm{C})$ algebra, which was the starting point of the derivation, but they were not adjusted to performing the most general quantum contractions, which is why we were then describing the latter in a bit cumbersome manner. 

There are the following classes in the de Sitter case (we redefine the parameters $\gamma_\pm := (\gamma \pm \bar\gamma)/2$ and, for brevity, denote $\tilde{\cal P}_\mu \equiv \Lambda^{-1/2} {\cal P}_\mu$):
\begin{align}\label{eq:32.01b}
r_I(\chi) &= \chi\, \big({\cal J}_2 + \tilde{\cal P}_1\big) \wedge \tilde{\cal P}_0\,, \nonumber\\
r_{II}(\chi,\varsigma) &= \frac{\chi}{2} \Big(\big({\cal J}_2 + \tilde{\cal P}_1\big) \wedge \tilde{\cal P}_0 + {\cal J}_0 \wedge \big({\cal J}_1 - \tilde{\cal P}_2\big)\Big) - \frac{\varsigma}{2} \big({\cal J}_2 + \tilde{\cal P}_1\big) \wedge \big({\cal J}_1 - \tilde{\cal P}_2\big)\,, \nonumber\\
r_{III}(\gamma_+,\gamma_-,\eta) &= \gamma_+ \big({\cal J}_1 \wedge \tilde{\cal P}_2 - {\cal J}_2 \wedge \tilde{\cal P}_1\big) + \gamma_- \big({\cal J}_1 \wedge {\cal J}_2 - \tilde{\cal P}_1 \wedge \tilde{\cal P}_2\big) + \frac{\eta}{2}\, {\cal J}_0 \wedge \tilde{\cal P}_0\,, \nonumber\\
r_{IV}(\gamma,\varsigma) &= \gamma\, \big({\cal J}_1 \wedge {\cal J}_2 - \tilde{\cal P}_1 \wedge \tilde{\cal P}_2 - {\cal J}_0 \wedge \tilde{\cal P}_0\big) - \frac{\varsigma}{2} \big({\cal J}_2 + \tilde{\cal P}_1\big) \wedge \big({\cal J}_1 - \tilde{\cal P}_2\big)\,.
\end{align}
In particular, $r_{III}$ describes a generalization of the time- or spacelike $\kappa$-deformation if $\gamma_- \neq 0$ or $\gamma_+ \neq 0$, respectively, and the (generalized) lightlike $\kappa$-deformation is given by $r_{II}$ with $\chi \neq 0$; certain special cases of $r_{III}$ and $r_{IV}$ are associated with four possible Drinfeld double structures on de Sitter algebra \cite{Ballesteros:2013dy} (cf. \cite{Kowalski:2020qs}). $r_I$ and $r_{II}$ satisfy the homogeneous Yang-Baxter equation, while for the remaining ones we have
\begin{align}\label{eq:60.01a}
[[r_{III},r_{III}]] &= -2 \big(\gamma_+^2 - \gamma_-^2\big) \big({\cal J}_0 \wedge {\cal J}_1 \wedge {\cal J}_2 - \tfrac{1}{2} \Lambda^{-1} \epsilon^{\mu\nu\sigma} {\cal J}_\mu \wedge {\cal P}_\nu \wedge {\cal P}_\sigma\big) \nonumber\\
&- 4\gamma_+ \gamma_- \Lambda^{-1/2} \big(\tfrac{1}{2} \epsilon^{\mu\nu\sigma} {\cal J}_\mu \wedge {\cal J}_\nu \wedge {\cal P}_\sigma - \Lambda^{-1} {\cal P}_0 \wedge {\cal P}_1 \wedge {\cal P}_2\big)\,, \nonumber\\
[[r_{IV},r_{IV}]] &= 2\gamma^2 \big({\cal J}_0 \wedge {\cal J}_1 \wedge {\cal J}_2 - \tfrac{1}{2} \Lambda^{-1} \epsilon^{\mu\nu\sigma} {\cal J}_\mu \wedge {\cal P}_\nu \wedge {\cal P}_\sigma\big)\,,
\end{align}
with $\tfrac{1}{2} \epsilon^{\mu\nu\sigma} {\cal J}_\mu \wedge {\cal P}_\nu \wedge {\cal P}_\sigma - \Lambda\, {\cal J}_0 \wedge {\cal J}_1 \wedge {\cal J}_2$ and ${\cal P}_0 \wedge {\cal P}_1 \wedge {\cal P}_2 - \Lambda\, \tfrac{1}{2} \epsilon^{\mu\nu\sigma} {\cal J}_\mu \wedge {\cal J}_\nu \wedge {\cal P}_\sigma$ being the two linearly independent invariants of the $\mathfrak{so}(3,1)$ algebra, which reduce to the invariants of $\mathfrak{iso}(2,1)$ in the limit $\Lambda \to 0$. 

In the anti-de Sitter case, the situation is more complicated. Namely, the $\mathfrak{o}(2,2)$ algebra has three decompositions into the direct sums of algebras: $\mathfrak{sl}(2;\mathbbm{R}) \oplus \bar{\mathfrak{sl}}(2;\mathbbm{R})$, $\mathfrak{su}(1,1) \oplus \bar{\mathfrak{su}}(1,1)$ and $\mathfrak{su}(1,1) \oplus \bar{\mathfrak{sl}}(2;\mathbbm{R})$, and each of them is associated with different $r$-matrices (see \cite{Borowiec:2017bs} or \cite{Kowalski:2020qs} for more details). In order to distinguish $r$-matrices coming from the latter two decompositions in the current paper, we will denote the ones of $\mathfrak{su}(1,1) \oplus \bar{\mathfrak{su}}(1,1)$ with a prime, and the ones of $\mathfrak{su}(1,1) \oplus \bar{\mathfrak{sl}}(2;\mathbbm{R})$ with a double prime. Then, the classification can be written as (redefining the parameters $\chi_\pm := (\chi \pm \bar\chi)/2$, $\gamma_\pm := (\gamma \pm \bar\gamma)/2$ and $\varrho_\pm := (\rho \pm \bar\chi)/2$, and denoting $\tilde{\cal P}_\mu \equiv |\Lambda|^{-1/2} {\cal P}_\mu$):
\begin{align}\label{eq:33.03a}
r_I(\chi) &= -\chi \big({\cal J}_0 - \tilde{\cal P}_2\big) \wedge \tilde{\cal P}_1\,, \nonumber\\
r_{II}(\chi_+,\chi_-,\varsigma) &= -\frac{\chi_+}{2} \Big(\big({\cal J}_0 - \tilde{\cal P}_2\big) \wedge \tilde{\cal P}_1 + {\cal J}_1 \wedge \big({\cal J}_2 - \tilde{\cal P}_0\big)\Big) \nonumber\\
&- \frac{\chi_-}{2} \Big(\big({\cal J}_2 - \tilde{\cal P}_0\big) \wedge \tilde{\cal P}_1 + {\cal J}_1 \wedge \big({\cal J}_0 - \tilde{\cal P}_2\big)\Big) - \frac{\varsigma}{2} \big({\cal J}_0 - \tilde{\cal P}_2\big) \wedge \big({\cal J}_2 - \tilde{\cal P}_0\big)\,, \nonumber\\
r_{III}(\gamma_+,\gamma_-,\eta) &= -\gamma_+ \big({\cal J}_0 \wedge \tilde{\cal P}_2 - {\cal J}_2 \wedge \tilde{\cal P}_0\big) + \gamma_- \big({\cal J}_0 \wedge {\cal J}_2 + \tilde{\cal P}_0 \wedge \tilde{\cal P}_2\big) - \frac{\eta}{2}\, {\cal J}_1 \wedge \tilde{\cal P}_1\,, \nonumber\\
r_{IV}(\gamma,\varsigma) &= \gamma\, \big({\cal J}_0 \wedge {\cal J}_2 + {\cal J}_1 \wedge \tilde{\cal P}_1 + \tilde{\cal P}_0 \wedge \tilde{\cal P}_2\big) - \frac{\varsigma}{2} \big({\cal J}_0 - \tilde{\cal P}_2\big) \wedge \big({\cal J}_2 - \tilde{\cal P}_0\big)\,, \nonumber\\
r_V(\gamma,\varrho_+,\varrho_-) &= \frac{\gamma}{2} \big({\cal J}_0 - \tilde{\cal P}_0\big) \wedge \big({\cal J}_2 - \tilde{\cal P}_2\big) + \frac{1}{2} \big(\varrho_+ {\cal J}_1 - \varrho_- \tilde{\cal P}_1\big) \wedge \big({\cal J}_0 - {\cal J}_2 + \tilde{\cal P}_0 - \tilde{\cal P}_2\big)
\end{align}
coming from $\mathfrak{sl}(2;\mathbbm{R}) \oplus \bar{\mathfrak{sl}}(2;\mathbbm{R})$;
\begin{align}\label{eq:35.03}
r_{III'}(\gamma_+,\gamma_-,\eta) = -\gamma_+ \big({\cal J}_1 \wedge \tilde{\cal P}_2 - {\cal J}_2 \wedge \tilde{\cal P}_1\big) + \gamma_- \big({\cal J}_1 \wedge {\cal J}_2 + \tilde{\cal P}_1 \wedge \tilde{\cal P}_2\big) + \frac{\eta}{2}\, {\cal J}_0 \wedge \tilde{\cal P}_0
\end{align}
coming from $\mathfrak{su}(1,1) \oplus \bar{\mathfrak{su}}(1,1)$; and
\begin{align}\label{eq:36.03}
r_{III''}(\gamma,\bar\gamma,\eta) &= \frac{\gamma}{2} \big({\cal J}_1 - \tilde{\cal P}_1\big) \wedge \big({\cal J}_2 - \tilde{\cal P}_2\big) - \frac{\bar\gamma}{2} \big({\cal J}_0 + \tilde{\cal P}_0\big) \wedge \big({\cal J}_2 + \tilde{\cal P}_2\big) \nonumber\\
&- \frac{\eta}{4} \big({\cal J}_0 - \tilde{\cal P}_0\big) \wedge \big({\cal J}_1 + \tilde{\cal P}_1\big)\,, \nonumber\\
r_{V''}(\gamma,\bar\chi,\rho) &= \frac{\gamma}{2} \big({\cal J}_1 - \tilde{\cal P}_1\big) \wedge \big({\cal J}_2 - \tilde{\cal P}_2\big) \nonumber\\
&+ \frac{1}{4} \Big(\bar\chi\, \big({\cal J}_1 + \tilde{\cal P}_1\big) + \rho\, \big({\cal J}_0 - \tilde{\cal P}_0\big)\Big) \wedge \big({\cal J}_0 - {\cal J}_2 + \tilde{\cal P}_0 - \tilde{\cal P}_2\big)
\end{align}
coming from $\mathfrak{su}(1,1) \oplus \bar{\mathfrak{sl}}(2;\mathbbm{R})$. $r_I$ and $r_{II}$ satisfy the homogeneous Yang-Baxter equation, while for the remaining classes we have
\begin{align}\label{eq:60.01c}
[[r_{III},r_{III}]] &= -2 \big(\gamma_+^2 + \gamma_-^2\big) \big({\cal J}_0 \wedge {\cal J}_1 \wedge {\cal J}_2 - \tfrac{1}{2} \Lambda^{-1} \epsilon^{\mu\nu\sigma} {\cal J}_\mu \wedge {\cal P}_\nu \wedge {\cal P}_\sigma\big) \nonumber\\
&- 4\gamma_+ \gamma_- |\Lambda|^{-1/2} \big(\tfrac{1}{2} \epsilon^{\mu\nu\sigma} {\cal J}_\mu \wedge {\cal J}_\nu \wedge {\cal P}_\sigma - \Lambda^{-1} {\cal P}_0 \wedge {\cal P}_1 \wedge {\cal P}_2\big)\,, \nonumber\\
[[r_{IV},r_{IV}]] &= -2\gamma^2 \big({\cal J}_0 \wedge {\cal J}_1 \wedge {\cal J}_2 - \tfrac{1}{2} \Lambda^{-1} \epsilon^{\mu\nu\sigma} {\cal J}_\mu \wedge {\cal P}_\nu \wedge {\cal P}_\sigma\big)\,, \nonumber\\
[[r_V,r_V]] &= -\gamma^2 \big({\cal J}_0 \wedge {\cal J}_1 \wedge {\cal J}_2 - \tfrac{1}{2} \Lambda^{-1} \epsilon^{\mu\nu\sigma} {\cal J}_\mu \wedge {\cal P}_\nu \wedge {\cal P}_\sigma \nonumber\\
&- \tfrac{1}{2} |\Lambda|^{-1/2} \epsilon^{\mu\nu\sigma} {\cal J}_\mu \wedge {\cal J}_\nu \wedge {\cal P}_\sigma + |\Lambda|^{-3/2} {\cal P}_0 \wedge {\cal P}_1 \wedge {\cal P}_2\big)
\end{align}
and
\begin{align}\label{eq:60.01e}
[[r_{III'},r_{III'}]] &= 2 \big(\gamma_+^2 + \gamma_-^2\big) \big({\cal J}_0 \wedge {\cal J}_1 \wedge {\cal J}_2 - \tfrac{1}{2} \Lambda^{-1} \epsilon^{\mu\nu\sigma} {\cal J}_\mu \wedge {\cal P}_\nu \wedge {\cal P}_\sigma\big) \nonumber\\
&+ 4\gamma_+ \gamma_- |\Lambda|^{-1/2} \big(\tfrac{1}{2} \epsilon^{\mu\nu\sigma} {\cal J}_\mu \wedge {\cal J}_\nu \wedge {\cal P}_\sigma - \Lambda^{-1} {\cal P}_0 \wedge {\cal P}_1 \wedge {\cal P}_2\big)\,, \nonumber\\
[[r_{III''},r_{III''}]] &= \big(\gamma^2 - \bar\gamma^2\big) \big({\cal J}_0 \wedge {\cal J}_1 \wedge {\cal J}_2 - \tfrac{1}{2} \Lambda^{-1} \epsilon^{\mu\nu\sigma} {\cal J}_\mu \wedge {\cal P}_\nu \wedge {\cal P}_\sigma\big) \nonumber\\
&+ \big(\gamma^2 + \bar\gamma^2\big) \big(\tfrac{1}{2} |\Lambda|^{-1/2} \epsilon^{\mu\nu\sigma} {\cal J}_\mu \wedge {\cal J}_\nu \wedge {\cal P}_\sigma + |\Lambda|^{-3/2} {\cal P}_0 \wedge {\cal P}_1 \wedge {\cal P}_2\big)\,, \nonumber\\
[[r_{V''},r_{V''}]] &= \gamma^2 \big({\cal J}_0 \wedge {\cal J}_1 \wedge {\cal J}_2 - \tfrac{1}{2} \Lambda \epsilon^{\mu\nu\sigma} {\cal J}_\mu \wedge {\cal P}_\nu \wedge {\cal P}_\sigma \nonumber\\
&- \tfrac{1}{2} |\Lambda|^{-1/2} \epsilon^{\mu\nu\sigma} {\cal J}_\mu \wedge {\cal J}_\nu \wedge {\cal P}_\sigma - |\Lambda|^{-3/2} {\cal P}_0 \wedge {\cal P}_1 \wedge {\cal P}_2\big)\,,
\end{align}
where the two invariants of the $\mathfrak{so}(2,2)$ algebra have the same form as the ones of $\mathfrak{so}(3,1)$ in (\ref{eq:60.01a}). In contrast to the de Sitter case, $r_{III}$ contains two copies of the spacelike $\kappa$-deformation, parametrized by $\gamma_+$ and $\gamma_-$ (which differ by an automorphism), and the corresponding terms of $r_{III'}$ are two copies of the timelike $\kappa$-deformation; the lightlike $\kappa$-deformation, parametrized by $\chi_-$, is contained in $r_{II}$. Certain special cases of $r_{III}$ and $r_{IV}$ are associated with three possible Drinfeld double structures on anti-de Sitter algebra \cite{Ballesteros:2013dy} (cf. \cite{Kowalski:2020qs}). 

\begin{figure}[h]
\centering
\includegraphics{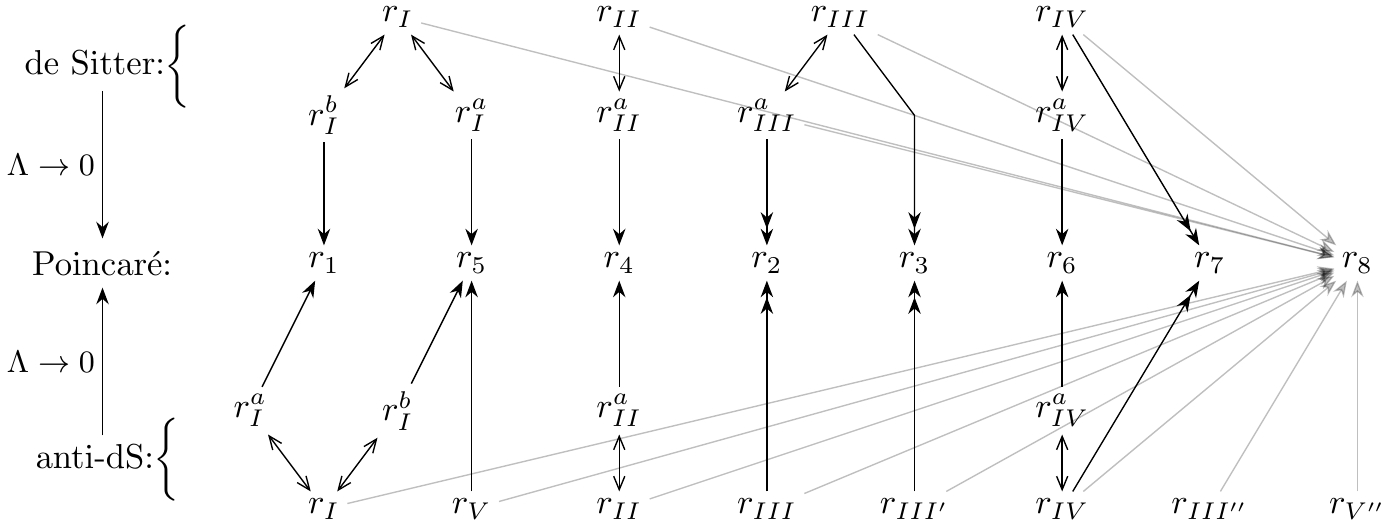}
\caption{\label{fig:1} Quantum ($\Lambda \to 0$) contractions relating all $r$-matrix classes for (anti-)de Sitter and Poincar\'{e} algebras; a two-headed arrow means that a given contraction recovers the full class; double arrows denote automorphisms of a given algebra; arrows leading to $r_8$ are lightened to make the diagram more legible.}
\end{figure}

It was shown by us in \cite{Kowalski:2020qs} that all of the $r$-matrix classes of Poincar\'{e} algebra (\ref{eq:60.01b}) can be recovered by the procedure of quantum contractions in the limit $\Lambda \to 0$ of $r$-matrices associated with de Sitter or anti-de Sitter algebras (given in (\ref{eq:32.01b}) and (\ref{eq:33.03a}-\ref{eq:36.03})). However, some of the terms proportional to the $\theta_{\mu\nu}$ parameters, as well the $\gamma$ term in $r_1$, turn out to be missing, i.e. not all contractions are surjective. We present these relations between the classes corresponding to $\Lambda > 0$, $\Lambda < 0$ and $\Lambda = 0$ in Fig.~\ref{fig:1}, depicted on a diagram that was not included in our previous paper.\footnote{We also overlooked in \cite{Kowalski:2020qs} that the contractions from the adS $r_{II}^a$ to $r_4$ and from $r_V$ to $r_5$ are not surjective, as can be shown using appropriate automorphisms of $\mathfrak{iso}(2,1)$.}

\section{Carrollian and Galilean (quantum) contractions}\label{sec:3.0}

\subsection{Contractions of Poincar\'{e} and (a)dS algebras}\label{sec:3.1}

As a more convenient basis to introduce the Carrollian/Galilean kinematical algebras, let us choose
\begin{align}\label{eq:10x}
J_0 := -{\cal J}_0\,, \qquad K_{1/2} := -{\cal J}_{1/2}\,, \qquad P_0 := {\cal P}_0\,, \qquad P_{1/2} := \mp {\cal P}_{2/1}\,.
\end{align}
The brackets of Poincar\'{e} algebra (\ref{eq:32.03}) then become
\begin{align}\label{eq:10}
[J_0,K_a] & = \epsilon_a^{\ b} K_b\,, & [K_1,K_2] & = -J_0\,, & 
[J_0,P_a] & = \epsilon_a^{\ b} P_b\,, & [J_0,P_0] & = 0\,, \nonumber\\ 
[K_a,P_b] & = \delta_{ab} P_0\,, & [K_a,P_0] & = P_a\,, & [P_1,P_2] & = 0\,, & [P_0,P_a] & = 0
\end{align}
($a = 1,2$ and indices are raised with Euclidean metric). If we denote $J := J_0$, $T_a := P_a$ and define the rescaled generators
\begin{align}\label{eq:11y}
Q_a := c\, K_a\,, \qquad T_0 := c\, P_0\,,
\end{align}
it allows us to perform a contraction of Poincar\'{e} (Lie) algebra by taking the limit $c \to 0$ of (\ref{eq:10}) to obtain the brackets of 2+1-dimensional Carroll algebra:
\begin{align}\label{eq:11}
[J,Q_a] & = \epsilon_a^{\ b} Q_b\,, & [Q_1,Q_2] & = 0\,, & 
[J,T_a] & = \epsilon_a^{\ b} T_b\,, & [J,T_0] & = 0\,, \nonumber\\ 
[Q_a,T_b] & = \delta_{ab} T_0\,, & [Q_a,T_0] & = 0\,, & [T_1,T_2] & = 0\,, & [T_0,T_a] & = 0\,.
\end{align}
(Technically, the generators $Q_a$ and $T_0$ in this limit should be denoted using other symbols, since they are no longer the rescaled generators of Poincar\'{e} algebra but generators of Carroll algebra. However, such a simplification of notation should not lead to confusion.) One of the interesting properties of this algebra, which follows from what we discussed in Introduction, is that it can be embedded as a subalgebra of Poincar\'{e} algebra in 3+1 dimensions, see Appendix~\ref{sec:A}. On the other hand, Carroll algebra possesses automorphisms without counterparts in the Poincar\'{e} case of the same dimension, which mix generators of boosts and spatial translations:
\begin{align}\label{eq:11z}
Q_a \rightarrow \alpha\, Q_a + \beta\, T_a\,, \quad T_a \rightarrow \alpha\, T_a - \beta\, Q_a\,,
\end{align}
where $\alpha^2 + \beta^2 = 1$. Both facts turn out to be relevant from the perspective of quantum deformations. 

On the other hand, introducing the rescaled generators
\begin{align}\label{eq:11a}
Q_a := c^{-1} K_a\,, \qquad T_a := c^{-1} P_a
\end{align}
and denoting $J := J_0$, $T_0 := P_0$, we may perform another contraction of Poincar\'{e} algebra -- take the limit $c \to \infty$ of (\ref{eq:10}) to obtain the brackets of 2+1-dimensional Galilei algebra:
\begin{align}\label{eq:11c}
[J,Q_a] & = \epsilon_a^{\ b} Q_b\,, & [Q_1,Q_2] & = 0\,, & 
[J,T_a] & = \epsilon_a^{\ b} T_b\,, & [J,T_0] & = 0\,, \nonumber\\ 
[Q_a,T_b] & = 0\,, & [Q_a,T_0] & = T_a\,, & [T_1,T_2] & = 0\,, & [T_0,T_a] & = 0\,.
\end{align}
(The notation is kept the same as for Carroll algebra but it will be clear from the context which algebra we consider at a given moment.) Calculating the cobrackets determined by all possible products of generators, we find that the algebra has the following antisymmetric split-Casimir
\begin{align}\label{eq:11cx}
{\cal C}_{\rm s1} := Q_1 \wedge T_1 + Q_2 \wedge T_2\,.
\end{align}
As we mentioned while defining the cobracket (\ref{eq:10.00}) for a coboundary Lie bialgebra, such a tensor gives a trivial contribution to a classical $r$-matrix and hence we will always drop it in the Galilean contractions. This is in line with the observation made in \cite{Ballesteros:2020ts} in the case of 3+1 spacetime dimensions, where a $r$-matrix for Galilei algebra (as well as for (anti-)de Sitter-Galilei algebra -- see below) that consists only of the terms analogous to (\ref{eq:11cx}) was shown to lead to the vanishing cobrackets. 

The change of basis (\ref{eq:10x}) applied to the brackets of (anti-)de Sitter algebra (\ref{eq:32.03a}) gives
\begin{align}\label{eq:10a}
[J_0,K_a] & = \epsilon_a^{\ b} K_b\,, & [K_1,K_2] & = -J_0\,, & 
[J_0,P_a] & = \epsilon_a^{\ b} P_b\,, & [J_0,P_0] & = 0\,, \nonumber\\ 
[K_a,P_b] & = \delta_{ab} P_0\,, & [K_a,P_0] & = P_a\,, & [P_1,P_2] & = \Lambda\, J_0\,, & [P_0,P_a] & = -\Lambda\, K_a\,.
\end{align}
If we redefine the generators as in (\ref{eq:11y}) and take the Carroll limit ($c \to 0$), it leads to (anti-)de Sitter-Carroll algebra, also known as the ``para-Euclidean'' in the case of $\Lambda > 0$, and ``para-Poincar\'{e}'' in the case of $\Lambda < 0$:
\begin{align}\label{eq:11d}
[J,Q_a] & = \epsilon_a^{\ b} Q_b\,, & [Q_1,Q_2] & = 0\,, & 
[J,T_a] & = \epsilon_a^{\ b} T_b\,, & [J,T_0] & = 0\,, \nonumber\\ 
[Q_a,T_b] & = \delta_{ab} T_0\,, & [Q_a,T_0] & = 0\,, & [T_1,T_2] & = \Lambda\, J\,, & [T_a,T_0] & =  \Lambda\, Q_a\,.
\end{align}
In the limit $\Lambda \to 0$, the brackets (\ref{eq:11d}) reduce to Carroll algebra (\ref{eq:11}). 

Analogously, rescaling the generators of (\ref{eq:10a}) as in (\ref{eq:11a}) and taking the Galilei limit ($c \to \infty$), we obtain (anti-)de Sitter-Galilei algebra, also known as the ``expanding Newton-Hooke'' in the case of $\Lambda > 0$ and ``oscillating Newton-Hooke'' in the case of $\Lambda < 0$:
\begin{align}\label{eq:11e}
[J,Q_a] & = \epsilon_a^{\ b} Q_b\,, & [Q_1,Q_2] & = 0\,, & 
[J,T_a] & = \epsilon_a^{\ b} T_b\,, & [J,T_0] & = 0\,, \nonumber\\ 
[Q_a,T_b] & = 0\,, & [Q_a,T_0] & = T_a\,, & [T_1,T_2] & = 0\,, & [T_a,T_0] & = \Lambda\, Q_a\,.
\end{align}
Galilei algebra (\ref{eq:11c}) is recovered in the limit $\Lambda \to 0$. Each of the above Galilean algebras with $\Lambda \neq 0$ has an antisymmetric split-Casimir of the same form (\ref{eq:11cx}) as for Galilei algebra, as well as another one
\begin{align}\label{eq:11ex}
{\cal C}_{\rm s2} := Q_1 \wedge Q_2 - \Lambda^{-1} T_1 \wedge T_2\,.
\end{align}
What will be also relevant to our investigations is that both algebras possess automorphisms mixing generators of boosts and spatial translations:
\begin{align}\label{eq:11ey}
Q_{1/2} \mapsto \alpha\, Q_{1/2} \pm \beta \Lambda^{-1/2} T_{2/1}\,, \quad T_{1/2} \mapsto \alpha\, T_{1/2} \pm \beta \Lambda^{1/2} Q_{2/1}\,,
\end{align}
where $\alpha^2 + \beta^2 = 1$, for $\Lambda > 0$ and $Q_a \mapsto |\Lambda|^{-1/2} T_a$, $T_a \mapsto -|\Lambda|^{1/2} Q_a$ for $\Lambda < 0$. 

\subsection{Quantum contractions of $r$-matrices for Poincar\'{e} algebra}\label{sec:3.2}

We will first calculate the Carrollian and Galilean quantum contractions of the $r$-matrices (\ref{eq:60.01}). The word ``quantum'' in this context does not mean that the procedure is in some sense quantum but it refers to the (quantum) deformation parameters (which control the quantization of a Lie bialgebra, leading to a Hopf algebra). Namely, performing (quantum) contractions of $r$-matrices involves also the appropriate rescalings of their parameters, so that the contraction limits are well-defined (cf. \cite{Kowalski:2020qs} for quantum contractions in the limit of $\Lambda \to 0$). In the Carrollian case, a given parameter $q$ usually needs to be rescaled either to $\tilde q := q/c$ or to $\hat q := q/c^2$. As the result, we obtain the following quantum contraction limits:
\begin{align}\label{eq:60.01ba}
r_{C2}(\hat\gamma,\tilde\eta;\{\theta_{\mu\nu}\}) &= \hat\gamma\, Q_2 \wedge T_0 - \tilde\eta\, Q_1 \wedge T_2 + r_{C8}(\{\theta_{\mu\nu}\})\,, \qquad \hat\gamma \neq 0 \vee \tilde\eta \neq 0\,, \nonumber\\
r_{C3}(\tilde\gamma,\tilde\eta;\{\theta_{\mu\nu}\}) &= \tilde\gamma\, (Q_1 \wedge T_1 + Q_2 \wedge T_2) - \tilde\eta\, J \wedge T_0 + r_{C8}(\{\theta_{\mu\nu}\})\,, \qquad \tilde\gamma \neq 0 \vee \tilde\eta \neq 0\,, \nonumber\\
r_{C6}(\hat\gamma,\tilde\gamma;\{\theta_{\mu\nu}\}) &= \hat\gamma\, Q_2 \wedge T_0 - \tilde\gamma\, (J \wedge T_0 + Q_1 \wedge T_2 - Q_2 \wedge T_1) + r_{C8}(\{\theta_{\mu\nu}\}) \nonumber\\
&\cong -\tilde\gamma\, (J \wedge T_0 + Q_1 \wedge T_2 - Q_2 \wedge T_1) + r_{C8}(\{\theta_{\mu\nu}\}) = r_{C6}(\tilde\gamma;\{\theta_{\mu\nu}\})\,, \nonumber\\
r_{C8}(\tilde\theta_{01},\theta_{12},\tilde\theta_{20}) &= \tilde\theta_{01} T_0 \wedge T_2 + \theta_{12} T_1 \wedge T_2 + \tilde\theta_{20} T_0 \wedge T_1\,.
\end{align}
The most general contraction of $r_6$ required the appropriate splitting and rescaling of the parameter $\varsigma$, $\varsigma = (\varsigma - \gamma) + \gamma \equiv c^2 \hat\gamma + c\, \tilde\gamma$ (if $\varsigma = 0$, the contraction limit becomes $r_{C2}(\hat\gamma,\tilde\eta = 0)$ instead). This is consistent with the fact that (as mentioned below (\ref{eq:60.01})) $\gamma \neq 0$, hence $\tilde\gamma \neq 0$, while the value of $\hat\gamma$ is left unconstrained. However, the automorphism $J \mapsto J + \hat\gamma/(2\tilde\gamma)\, Q_2$, $T_1 \mapsto T_1 - \hat\gamma/(2\tilde\gamma)\, T_0$ allows us to subsequently simplify the form of $r_6$ and get rid of $\hat\gamma$, which turns out to be redundant. 

The list (\ref{eq:60.01ba}) does not include the classes of $r$-matrices that ceased to be independent, i.e. became subsumed into other ones. Firstly, we note $r_{C5}(\hat\chi) = r_{C4}(\hat\chi,\hat\varsigma = 0)$. Secondly, acting with an appropriate automorphism (describing a rotation of the spatial axes) and since one of the parameters is again redundant, it can be shown that
\begin{align}\label{eq:60.01bx}
r_{C4}(\hat\chi,\hat\varsigma) = -\left(\hat\chi\, Q_1 + \hat\varsigma\, Q_2\right) \wedge T_0 \cong r_{C2}({\rm sgn}\hat\chi \sqrt{\hat\chi^2 + \hat\varsigma^2},\tilde\eta = 0)\,.
\end{align}
Thirdly, a Carroll algebra automorphism (\ref{eq:11z}) with $\alpha = 0$, $\beta = 1$ imposes the relations
\begin{align}\label{eq:60.01bb}
r_{C7}(\tilde\gamma) &= -\tilde\gamma\, (J \wedge T_0 - Q_1 \wedge T_2 + Q_2 \wedge T_1) \cong r_{C6}(\tilde\gamma;\{\theta_{\mu\nu} = 0\})\,, \nonumber\\
r_{C1}(\hat\chi,\tilde\gamma \neq 0) &= \hat\chi\, Q_1 \wedge Q_2 - \tilde\gamma\, (J \wedge T_0 - Q_1 \wedge T_2 + Q_2 \wedge T_1)\,, \nonumber\\
&\cong r_{C6}(\tilde\gamma;\theta_{01} = 0,\hat\chi,\theta_{20} = 0)\,, \nonumber\\
r_{C1}(\hat\chi,\tilde\gamma = 0) &\cong r_{C8}(\tilde\theta_{01} = 0,\hat\chi,\tilde\theta_{20} = 0)\,, \nonumber\\
r_{C8}(\tilde\theta_{01},\theta_{12},\tilde\theta_{20}) &\cong \theta_{12} Q_1 \wedge Q_2 + (\tilde\theta_{20} Q_1 + \tilde\theta_{01} Q_2) \wedge T_0\,. 
\end{align}
In order to remove the overlapping that now appears between $r_{C2}$ and $r_{C8}$, let us assume (in addition to the conditions given below (\ref{eq:60.01})) that $\theta_{12} \neq 0$ in $r_{C8}$. 

We also calculate (either explicitly or just by performing the Carrollian quantum contractions of the RHS-s of (\ref{eq:60.02})) that the $r$-matrices $r_{C2}$ and $r_{C8}$ satisfy the homogeneous Yang-Baxter equation, while for the remaining ones the equations remain inhomogeneous:
\begin{align}\label{eq:60.02b}
[[r_{C3},r_{C3}]] &= 2\tilde\gamma^2\, Q^a \wedge T_a \wedge T_0 + 4\tilde\gamma\, \theta_{12}\, T_0 \wedge T_1 \wedge T_2 \nonumber\\
[[r_{C6},r_{C6}]] &= -2\tilde\gamma^2\, Q^a \wedge T_a \wedge T_0\,.
\end{align}
As expected, $Q^a \wedge T_a \wedge T_0$ and $T_0 \wedge T_1 \wedge T_2$ are invariants of Carroll algebra but its peculiar feature is that automorphisms (\ref{eq:11z}) allow to transform the second invariant into a linear combination including two additional invariants,
\begin{align}\label{eq:60.02bx}
T_0 \wedge T_1 \wedge T_2 \cong \alpha^2 T_0 \wedge T_1 \wedge T_2 + \beta^2 T_0 \wedge Q_1 \wedge Q_2 + \alpha \beta\, T_0 \wedge (Q_1 \wedge T_2 - Q_2 \wedge T_1)\,.
\end{align}
Another thing to note is a reduction of $r_2$, describing the (twisted) $\kappa$-deformation in a spatial direction, to a triangular $r$-matrix $r_{C2}$, while $r_3$, describing the analogous deformation in the time direction, does not simplify in the same way. 

In the Galilean case, relevant rescalings of a given deformation parameter $q$ are $\tilde q := c\, q$ and $\hat q := c^2 q$. After the $c \to \infty$ contractions of (\ref{eq:60.01}) are performed, our results are simplified by dropping the (irrelevant) terms proportional to the antisymmetric split-Casimir (\ref{eq:11cx}), i.e. it makes some $r$-matrices depend on fewer parameters than their Poincar\'{e} progenitors. Therefore, we obtain the following quantum contraction limits:
\begin{align}\label{eq:60.01bc}
r_{G1}(\hat\chi,\hat\gamma) &= \hat\chi\, Q_1 \wedge Q_2 + \hat\gamma\, (Q_1 \wedge T_2 - Q_2 \wedge T_1)\,, \qquad \hat\chi \neq 0\,, \nonumber\\
r_{G2}(\tilde\gamma,\hat\eta;\{\theta_{\mu\nu}\}) &= \tilde\gamma\, (J \wedge T_1 + Q_2 \wedge T_0) - \hat\eta\, Q_1 \wedge T_2 + r_{G8}(\{\theta_{\mu\nu}\})\,, \qquad \tilde\gamma \neq 0\,, \nonumber\\
r_{G3}(\eta;\{\theta_{\mu\nu}\}) &= 
-\eta\, J \wedge T_0 + r_{G8}(\{\theta_{\mu\nu}\})\,, 
\nonumber\\
r_{G5}(\hat\chi;\{\theta_{\mu\nu}\}) &= \hat\chi\, Q_1 \wedge T_1 + r_{G8}(\{\theta_{\mu\nu}\})\,, 
\nonumber\\
r_{G6}(\hat\gamma,\hat\varsigma;\{\theta_{\mu\nu}\}) &= -\hat\gamma\, Q_1 \wedge T_2 + \hat\varsigma\, Q_2 \wedge T_1 + r_{G8}(\{\theta_{\mu\nu}\})\,, \qquad \hat\gamma \neq 0 \vee \hat\varsigma \neq 0\,, \nonumber\\
r_{G8}(\tilde\theta_{01},\hat\theta_{12},\tilde\theta_{20}) &= \tilde\theta_{01} T_0 \wedge T_2 + \hat\theta_{12} T_1 \wedge T_2 + \tilde\theta_{20} T_0 \wedge T_1\,.
\end{align}
The list is given without the $r$-matrix classes that become subsumed into other ones, to wit $r_{G4}(\hat\varsigma) = r_{G6}(\hat\gamma = 0,\hat\varsigma)$ and $r_{G7}(\hat\gamma) = -r_{G6}(\hat\gamma,\hat\gamma;\{\theta_{\mu\nu} = 0\})$. Moreover, we modified restrictions on the parameters of $r_{G2}$ and $r_{G6}$ with respect to the ones given below (\ref{eq:60.01}), so that the classes contain all possible cases but remain disjoint. 

\begin{figure}[h]
\centering
\includegraphics{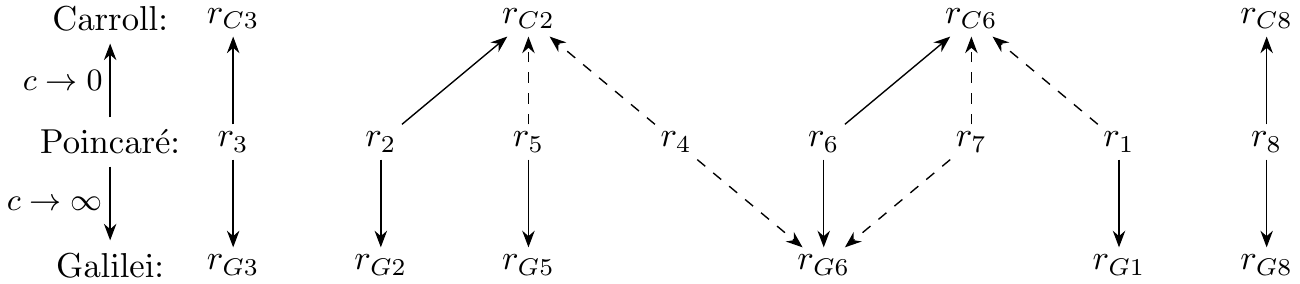}
\caption{\label{fig:4} Quantum ($c \to 0$ and $c \to \infty$) contractions relating all $r$-matrix classes for Poincar\'{e} algebra, and those obtained for Carroll and Galilei algebras; a dashed line means that a given contraction leads to a subclass of a larger class.}
\end{figure}

Interestingly, we find (either by an explicit calculation or by performing the Galilean quantum contractions of the RHS-s of (\ref{eq:60.02})) that the above $r$-matrices satisfy the homogeneous Yang-Baxter equation in all cases, apart from
\begin{align}\label{eq:60.02c}
[[r_{G2},r_{G2}]] = -2\tilde\gamma^2 (-J \wedge T_1 \wedge T_2 + Q^a \wedge T_a \wedge T_0) + 4\tilde\gamma\, \tilde\theta_{20}\, T_0 \wedge T_1 \wedge T_2\,.
\end{align}
$-J \wedge T_1 \wedge T_2 + Q^a \wedge T_a \wedge T_0$ and $T_0 \wedge T_1 \wedge T_2$ are the two invariants of Galilei algebra. Conversely to the Carrollian case, it is the $r$-matrix $r_3$, describing the (twisted) $\kappa$-deformation in the time direction, that now reduces to a triangular $r$-matrix $r_{G3}$. We will comment on this fact in conclusions in Sec.~\ref{sec:6.0}. The results of this Subsection are illustrated by a diagram in Fig.~\ref{fig:4}. For clarity of the figure, we do not show that $r_1(\gamma = 0)$ contracts to $r_{C8}$, while $r_6(\varsigma = 0)$ and $r_8(\theta_{12} = 0)$ contract to $r_{C2}$, as well as $r_2(\gamma = 0)$ contracts to $r_{G6}$, while $r_3(\eta = 0)$ and $r_4(\varsigma = 0)$ contract to $r_{G8}$.

\section{(a)dS-Carroll algebras and their quantum deformations}\label{sec:4.0}

\subsection{The complete classification of adSC and dSC $r$-matrices}\label{sec:4.1}

Every kinematical algebra considered in this paper generates the group of symmetries of a different homogeneous space, i.e. a different spacetime. However, the algebras alone (irrespective of their physical interpretation) may in some cases be equivalent. Indeed, adS-Carroll algebra is isomorphic to Poincar\'{e} algebra (this is true in any number of dimensions and explains the name ``para-Poincar\'{e}''), since the brackets (\ref{eq:10}) can be transformed into (\ref{eq:11d}) with $\Lambda < 0$ via a map
\begin{align}\label{eq:12}
K_a \mapsto |\Lambda|^{-1/2} T_a\,, \quad P_a \mapsto -|\Lambda|^{1/2} Q_a\,, \quad J_0 \mapsto J\,, \quad P_0 \mapsto T_0
\end{align}
(or instead with $K_a \mapsto -|\Lambda|^{-1/2} T_a$, $P_a \mapsto |\Lambda|^{1/2} Q_a$). It straightforwardly follows that the complete classification of $r$-matrices for adS-Carroll algebra is obtained by expressing the classification for Poincar\'{e} algebra (\ref{eq:60.01b}) in the basis (\ref{eq:10x}) and acting on it with (\ref{eq:12}). Moreover, as in the Poincar\'{e} case, the classification actually contains all possible Hopf-algebraic deformations of the algebra. To our knowledge, these simple facts have not been noticed before. 

After applying the above isomorphism, we rescale all $\theta$ parameters by $|\Lambda|$ and the remaining ones, except $\chi$ in $r_1$, by $|\Lambda|^{1/2}$. In such a way, we obtain (for brevity, let us denote $\tilde T_\mu \equiv |\Lambda|^{-1/2} T_\mu$):
\begin{align}\label{eq:60.01x}
r_{1'}(\chi,\gamma) &= \chi\, (J + \tilde T_1) \wedge \tilde T_2 - \gamma (J \wedge \tilde T_0 + Q_1 \wedge \tilde T_2 - Q_2 \wedge \tilde T_1)\,, \nonumber\\
r_{2'}(\gamma,\eta,\theta_{20}) &= -\gamma (J \wedge Q_1 + \tilde T_0 \wedge \tilde T_2) - \eta\, Q_2 \wedge \tilde T_1 - \theta_{20} Q_1 \wedge \tilde T_0\,, \nonumber\\
r_{3'}(\gamma,\eta,\theta_{12}) &= \gamma (Q_1 \wedge \tilde T_1 + Q_2 \wedge \tilde T_2) - \eta\, J \wedge \tilde T_0 + \theta_{12} Q_1 \wedge Q_2\,, \nonumber\\
r_{4'}(\chi,\varsigma,\theta) &= \chi \big((Q_1 + \tilde T_0) \wedge \tilde T_1 + Q_2 \wedge (J + \tilde T_2)\big) - \varsigma\, (J + \tilde T_2) \wedge (Q_1 + \tilde T_0) \nonumber\\
&- \theta\, Q_2 \wedge (Q_1 - \tilde T_0)\,, \nonumber\\
r_{5'}(\chi,\theta_{01},\theta_{12}) &= \chi\, (Q_1 + \tilde T_0) \wedge \tilde T_1 + \theta_{12} Q_1 \wedge Q_2 + \theta_{01} Q_2 \wedge \tilde T_0\,, \nonumber\\
r_{6'}(\gamma,\varsigma,\theta,\theta_{20}) &= -\gamma (J \wedge Q_1 + \tilde T_0 \wedge \tilde T_2 + Q_2 \wedge \tilde T_1) - \varsigma\, (J + \tilde T_2) \wedge (Q_1 + \tilde T_0) \nonumber\\
&- \theta\, Q_2 \wedge (Q_1 - \tilde T_0) - \theta_{20} Q_1 \wedge \tilde T_0\,, \nonumber\\
r_{7'}(\gamma) &= -\gamma (J \wedge \tilde T_0 + Q_1 \wedge \tilde T_2 - Q_2 \wedge \tilde T_1)\,, \nonumber\\
r_{8'}(\theta_{01},\theta_{12},\theta_{20}) &= \theta_{12} Q_1 \wedge Q_2 + \theta_{01} Q_2 \wedge \tilde T_0 + \theta_{20} Q_1 \wedge \tilde T_0\,.
\end{align}
The restrictions on the values of parameters are also inherited from (\ref{eq:60.01b}), i.e. $\gamma \neq 0 \vee \eta \neq 0$ and $\chi \neq 0 \vee \varsigma \neq 0$, as well as $\gamma \neq \eta$ (in $r_{2'}$), $\chi \neq 0$ (in $r_{1'}$ and $r_{5'}$) and $\gamma \neq 0$ (in $r_{6'}$ and $r_{7'}$). Similarly, we find that the corresponding Yang-Baxter equations become
\begin{align}\label{eq:60.02x}
[[r_{1'},r_{1'}]] = [[r_{7'},r_{7'}]] &= 2\gamma^2 \big(J \wedge Q_1 \wedge Q_2 + \Lambda^{-1} Q^a \wedge T_a \wedge T_0\big)\,, \nonumber\\
[[r_{2'},r_{2'}]] = [[r_{6'},r_{6'}]] &= 2\gamma^2 \big(J \wedge Q_1 \wedge Q_2 + \Lambda^{-1} Q^a \wedge T_a \wedge T_0\big) + 4\gamma\, \theta_{20} |\Lambda|^{-1/2} Q_1 \wedge Q_2 \wedge T_0\,, \nonumber\\
[[r_{3'},r_{3'}]] &= -2\gamma^2 \big(J \wedge Q_1 \wedge Q_2 + \Lambda^{-1} Q^a \wedge T_a \wedge T_0\big) - 4\gamma\, \theta_{12} |\Lambda|^{-1/2} Q_1 \wedge Q_2 \wedge T_0\,, \nonumber\\
[[r_{4'},r_{4'}]] &= 8\chi\, \theta\, |\Lambda|^{-1/2} Q_1 \wedge Q_2 \wedge T_0\,, \nonumber\\
[[r_{5'},r_{5'}]] &= 2\chi (\theta_{01} + \theta_{12}) |\Lambda|^{-1/2} Q_1 \wedge Q_2 \wedge T_0
\end{align}
and the homogeneous one for $r_{8'}$. 

A complementary observation (explaining the name ``para-Euclidean'') is that dS-Carroll algebra is isomorphic to Euclidean algebra, sometimes called the inhomogeneous Euclidean algebra. The 3-dimensional version of the latter is $\mathfrak{iso}(3) = \mathfrak{so}(3) \vartriangleright\!\!< \mathbbm{R}^3$ and if its brackets are written down in the manner analogous to (\ref{eq:10}),
\begin{align}\label{eq:10b}
[J_3,K_a] & = \epsilon_a^{\ b} K_b\,, & [K_1,K_2] & = J_3\,, & 
[J_3,P_a] & = \epsilon_a^{\ b} P_b\,, & [J_3,P_3] & = 0\,, \nonumber\\ 
[K_a,P_b] & = -\delta_{ab} P_3\,, & [K_a,P_3] & = P_a\,, & [P_1,P_2] & = 0\,, & [P_3,P_a] & = 0\,,
\end{align}
we easily notice a map
\begin{align}\label{eq:12a}
K_a \mapsto \Lambda^{-1/2} T_a\,, \quad P_a \mapsto \Lambda^{1/2} Q_a\,, \quad J_3 \mapsto J\,, \quad P_3 \mapsto T_0
\end{align}
that transforms them into (\ref{eq:11d}) with $\Lambda > 0$. 

Meanwhile, the complete classification of classical $r$-matrices for $\mathfrak{iso}(3)$, which actually correspond to all possible Hopf-algebraic deformations of this inhomogeneous orthogonal algebra, has also been given in \cite{Stachura:1998ps} and can be put in the form \cite{Kowalski:2020qs}:
\begin{align}\label{eq:A0.02}
r_1(\gamma,\eta;\theta_{12}) &= -\gamma\, \big(K_1 \wedge P_1 + K_2 \wedge P_2\big) + \eta\, J_3 \wedge P_3 + \theta_{12} P_1 \wedge P_2\,, \nonumber\\
r_2(\gamma) &= \gamma\, \big(K_1 \wedge P_2 - K_2 \wedge P_1 + J_3 \wedge P_3\big)\,, \nonumber\\
r_3(\theta_{01},\theta_{12},\theta_{20}) &= \theta_{01} P_3 \wedge P_2 + \theta_{12} P_1 \wedge P_2 + \theta_{20} P_3 \wedge P_1\,,
\end{align}
with the deformation parameters of $r_1$ restricted by a condition $\gamma \neq 0 \vee \eta \neq 0$. If we now apply the map (\ref{eq:12a}) to (\ref{eq:A0.02}), as well as rescale all $\theta$ parameters by $\Lambda$ and the remaining ones by $\Lambda^{1/2}$, it leads to the complete classification of $r$-matrices for dS-Carroll algebra:
\begin{align}\label{eq:A0.02a}
r_{1'}(\gamma,\eta;\theta_{12}) &= \gamma\, \big(Q_1 \wedge \tilde T_1 + Q_2 \wedge \tilde T_2\big) + \eta\, J \wedge \tilde T_0 + \theta_{12} Q_1 \wedge Q_2\,, \nonumber\\
r_{2'}(\gamma) &= \gamma\, \big(J \wedge \tilde T_0 + Q_1 \wedge \tilde T_2 - Q_2 \wedge \tilde T_1\big)\,, \nonumber\\
r_{3'}(\theta_{01},\theta_{12},\theta_{20}) &= \theta_{12} Q_1 \wedge Q_2 - \theta_{01} Q_2 \wedge \tilde T_0 - \theta_{20} Q_1 \wedge \tilde T_0
\end{align}
(where, again, $\gamma \neq 0 \vee \eta \neq 0$ in $r_{1'}$). We find that $r_{3'}$ solves the homogeneous Yang-Baxter equation, while in other cases the equations have the form
\begin{align}\label{eq:A0.02ax}
[[r_{1'},r_{1'}]] &= 2\gamma^2 \big(J \wedge Q_1 \wedge Q_2 + \Lambda^{-1} Q^a \wedge T_a \wedge T_0\big) - 4\gamma\, \theta_{12} |\Lambda|^{-1/2} Q_1 \wedge Q_2 \wedge T_0\,, \nonumber\\
[[r_{2'},r_{2'}]] &= -2\gamma^2 \big(J \wedge Q_1 \wedge Q_2 + \Lambda^{-1} Q^a \wedge T_a \wedge T_0\big)\,.
\end{align}
The next two Subsections will show to what extent the classifications (\ref{eq:60.01x}) and (\ref{eq:A0.02a}) can be recovered as Carrollian quantum contraction limits of $r$-matrices of (anti-)de Sitter algebra, which have been recalled in Subsec.~\ref{sec:2.2}.

\subsection{Carrollian quantum contractions in the de Sitter case}

Let us start here with the de Sitter $r$-matrices (\ref{eq:32.01b}) and perform the procedure of quantum $c \to 0$ contractions in the same manner it was done for deformations of Poincar\'{e} algebra in (\ref{eq:60.01ba}). As the result, we obtain the following independent $r$-matrix classes:
\begin{align}\label{eq:32.01c}
r_{CII}(\hat\chi,\hat\varsigma) &= \frac{\hat\varsigma}{2}\, Q_1 \wedge Q_2 - \frac{\hat\chi}{2}\, \Lambda^{-1/2} Q_2 \wedge T_0\,, \nonumber\\
r_{CIII}(\hat\gamma_-,\tilde\gamma_+,\tilde\eta) &= \hat\gamma_- Q_1 \wedge Q_2 + \tilde\gamma_+ \Lambda^{-1/2} \big(Q_1 \wedge T_1 + Q_2 \wedge T_2\big) - \frac{\tilde\eta}{2}\, \Lambda^{-1/2} J \wedge T_0\,, \nonumber\\
r_{CIV}(\hat\gamma,\tilde\gamma) &= \hat\gamma\, Q_1 \wedge Q_2 + \tilde\gamma\, \Lambda^{-1/2} \big(J \wedge T_0 + Q_1 \wedge T_2 - Q_2 \wedge T_1\big) \nonumber\\
&\cong \tilde\gamma\, \Lambda^{-1/2} \big(J \wedge T_0 + Q_1 \wedge T_2 - Q_2 \wedge T_1\big) = r_{CIV}(\tilde\gamma)\,.
\end{align}
The class $r_{CI}$ is not listed, since it becomes a subclass of $r_{CII}$, i.e. $r_{CI}(\hat\chi) = r_{CII}(2\hat\chi,\hat\varsigma = 0)$. Similarly to the Poincar\'{e} $r$-matrix $r_6$, the most general contraction of $r_{IV}$ was derived due to the appropriate splitting and rescaling of $\varsigma$, $\varsigma = (\varsigma + 2\gamma) - 2\gamma \equiv 2c^2 \hat\gamma - 2c\, \tilde\gamma$, and the result was subsequently simplified by an automorphism of de Sitter-Carroll algebra, $T_a \mapsto T_a - \hat\gamma/(2\tilde\gamma)\, Q_a$, revealing that the parameter $\hat\gamma$ is actually redundant. 
 
Our previous study \cite{Kowalski:2020qs} of quantum $\Lambda \to 0$ contractions shows that one should also take into account that each of the de Sitter $r$-matrices (\ref{eq:32.01b}) can be transformed via certain automorphisms of de Sitter algebra $\mathfrak{so}(3,1)$ so that it changes the form of their dependence on different types of the generators (i.e., rotation, boosts and time/spatial translations). In principle, this could lead to the Carrollian (quantum) contraction limits not equivalent to any of (\ref{eq:32.01c}). The relevant automorphisms are the ones acting like rotations of the third spatial axis in the embedding 3+1-dimensional space and we choose their representative:
\begin{align}\label{eq:32.04fx}
{\cal J}_{0/2} \mapsto \Lambda^{-1/2} {\cal P}_{1/0}\,, \quad {\cal J}_1 \mapsto -{\cal J}_2\,, \quad {\cal P}_{0/2} \mapsto -\Lambda^{1/2} {\cal J}_{1/0}\,, \quad {\cal P}_1 \mapsto -{\cal P}_2\,,
\end{align}
which transforms (\ref{eq:32.01b}) into
\begin{align}\label{eq:32.04f}
r_{II}^a(\chi,\varsigma) &= \frac{\chi}{2} \Big({\cal J}_1 \wedge \big(\tilde{\cal P}_0 - \tilde{\cal P}_2\big) - \big({\cal J}_0 - {\cal J}_2\big) \wedge \tilde{\cal P}_1\Big) + \frac{\varsigma}{2} \big({\cal J}_0 - {\cal J}_2\big) \wedge \big(\tilde{\cal P}_0 - \tilde{\cal P}_2\big)\,, \nonumber\\
r_{III}^a(\gamma_-,\gamma_+,\eta) &= \gamma_- \big({\cal J}_0 \wedge \tilde{\cal P}_2 - {\cal J}_2 \wedge \tilde{\cal P}_0\big) - \gamma_+ \big({\cal J}_0 \wedge {\cal J}_2 - \tilde{\cal P}_0 \wedge \tilde{\cal P}_2\big) + \frac{\eta}{2}\, {\cal J}_1 \wedge \tilde{\cal P}_1\,, \nonumber\\
r_{IV}^a(\gamma,\varsigma) &= \gamma \big({\cal J}_0 \wedge \tilde{\cal P}_2 - {\cal J}_2 \wedge \tilde{\cal P}_0 - {\cal J}_1 \wedge \tilde{\cal P}_1\big) + \frac{\varsigma}{2} \big({\cal J}_0 - {\cal J}_2\big) \wedge \big(\tilde{\cal P}_0 - \tilde{\cal P}_2\big)\,.
\end{align}
The above $r$-matrices have the Carrollian contraction limits
\begin{align}\label{eq:32.04fa}
r_{CIIa}(\hat\chi,\hat\varsigma) &= -\frac{1}{2}\, \Lambda^{-1/2} \big(\hat\chi\, Q_1 - \hat\varsigma\, Q_2\big) \wedge T_0\,, \nonumber\\
r_{CIIIa}(\hat\gamma_-,\tilde\gamma_+,\tilde\eta) &= \hat\gamma_- \Lambda^{-1/2} Q_2 \wedge T_0 - \tilde\gamma_+ \big(J \wedge Q_2 + \Lambda^{-1} T_0 \wedge T_1\big) - \frac{\tilde\eta}{2}\, \Lambda^{-1/2} Q_1 \wedge T_2\,, \nonumber\\
r_{CIVa}(\hat\gamma,\tilde\gamma) &= \hat\gamma\, \Lambda^{-1/2} Q_2 \wedge T_0 + \tilde\gamma\, \Lambda^{-1/2} \big(J \wedge T_0 + Q_1 \wedge T_2 - Q_2 \wedge T_1\big)\,.
\end{align}
However, as one can check, the $\mathfrak{so}(3,1)$ automorphism (\ref{eq:32.04fx}) is inherited by de Sitter-Carroll algebra, and acting with it (expressed in the basis (\ref{eq:11d})) on $r_{CII}$, $r_{CIII}$ and $r_{CIV}$, we find out that the latter are equivalent to $r_{CIIa}$, $r_{CIIIa}$ and $r_{CIVa}$, respectively. The derivation of the expressions (\ref{eq:32.04fa}) is still illuminating, since they provide the standard form of $r$-matrices describing light- and spacelike $\kappa$-deformations, see Sec.~\ref{sec:6.0}, as well as will allow us to recover additional cases of $r$-matrices via quantum $\Lambda \to 0$ contractions, see Fig.~\ref{fig:2}. Meanwhile, applying the same kind of automorphisms to $r_I$, we are led to the contraction limits that explicitly belong to the class $r_{CII}$. 

Finally, comparing the results of this Subsection (\ref{eq:32.01c}) with the complete classification for dS-Carroll algebra (\ref{eq:A0.02a}), we can immediately identify:
\begin{align}\label{eq:A0.02b}
r_{CIII}(\hat\gamma_-,\tilde\gamma_+,\tilde\eta) &\cong r_{1'}(\gamma = \tilde\gamma_+,\eta = -\tilde\eta/2;\theta_{12} = \hat\gamma_-)\,, \nonumber\\
r_{CIV}(\tilde\gamma) &\cong r_{2'}(\gamma = \tilde\gamma)\,, \nonumber\\
r_{CII}(\hat\chi,\hat\varsigma) &\cong r_{3'}(\theta_{01} = \hat\chi/2,\theta_{12} = \hat\varsigma/2,\theta_{20} = 0)\,.
\end{align}
In conclusion, quantum contractions allowed us to recover all $r$-matrices in full generality ($\theta_{20}$ in $r_{3'}$ can always be brought to zero by an automorphism). If the same contractions are applied to the Yang-Baxter equations (\ref{eq:60.01a}), it naturally leads to the equations (\ref{eq:A0.02ax}) satisfied by $r_{1'}$, $r_{2'}$ and $r_{3'}$, respectively, up to the above renaming of parameters.

\subsection{Carrollian quantum contractions in the anti-de Sitter case}

Performing such contractions of the anti-de Sitter $r$-matrices (\ref{eq:33.03a}-\ref{eq:36.03}) analogously to how it was done in (\ref{eq:60.01ba}) and in the previous Subsection, we obtain:
\begin{align}\label{eq:33.03aa}
r_{CI}(\chi) &= \chi\, \big(J - |\Lambda|^{-1/2} T_1\big) \wedge |\Lambda|^{-1/2} T_2\,, \nonumber\\
r_{CII}(\hat\chi_+,\tilde\chi_-,\tilde\varsigma) &= -\frac{\hat\chi_+}{2}\, Q_1 \wedge \big(Q_2 + |\Lambda|^{-1/2} T_0\big) \nonumber\\
&+ \frac{\tilde\chi_-}{2} \Big(\big(Q_2 + |\Lambda|^{-1/2} T_0\big) \wedge |\Lambda|^{-1/2} T_2 - Q_1 \wedge \big(J - |\Lambda|^{-1/2} T_1\big)\Big) \nonumber\\
&- \frac{\tilde\varsigma}{2} \big(J - |\Lambda|^{-1/2} T_1\big) \wedge \big(Q_2 + |\Lambda|^{-1/2} T_0\big) \nonumber\\
&\cong \frac{\tilde\chi_-}{2} \Big(\big(Q_2 + |\Lambda|^{-1/2} T_0\big) \wedge |\Lambda|^{-1/2} T_2 - Q_1 \wedge \big(J - |\Lambda|^{-1/2} T_1\big)\Big) \nonumber\\
&- \frac{\tilde\varsigma}{2} \big(J - |\Lambda|^{-1/2} T_1\big) \wedge \big(Q_2 + |\Lambda|^{-1/2} T_0\big) = r_{CII}(\tilde\chi_-,\tilde\varsigma)\,, \nonumber\\
r_{CV}(\hat\gamma,\hat\varrho_+,\tilde\varrho_-) &= -\frac{\hat\gamma}{2}\, |\Lambda|^{-1/2} Q_2 \wedge T_0 - \frac{1}{2} \big(\hat\varrho_+ Q_1 - \tilde\varrho_- |\Lambda|^{-1/2} T_2\big) \wedge \big(Q_2 + |\Lambda|^{-1/2} T_0\big) \nonumber\\
&\cong -\frac{1}{2} \big(\hat\varrho_+ Q_1 - \tilde\varrho_- |\Lambda|^{-1/2} T_2\big) \wedge \big(Q_2 + |\Lambda|^{-1/2} T_0\big) = r_{CV}(\hat\varrho_+,\tilde\varrho_-)
\end{align}
and
\begin{align}\label{eq:35.03a}
r_{CIV}(\tilde\gamma,\tilde\varsigma) &= \tilde\gamma\, \big(J \wedge Q_2 + \Lambda^{-1} T_0 \wedge T_1 - |\Lambda|^{-1/2} Q_1 \wedge T_2\big) \nonumber\\
&- \frac{\tilde\varsigma}{2} \big(J - |\Lambda|^{-1/2} T_1\big) \wedge \big(Q_2 + |\Lambda|^{-1/2} T_0\big)\,, \nonumber\\
r_{CIII}(\hat\gamma_+,\tilde\gamma_-,\tilde\eta) &= -\hat\gamma_+ |\Lambda|^{-1/2} Q_2 \wedge T_0 + \tilde\gamma_- \big(J \wedge Q_2 + \Lambda^{-1} T_0 \wedge T_1\big) + \frac{\tilde\eta}{2}\, |\Lambda|^{-1/2} Q_1 \wedge T_2\,, \nonumber\\
r_{CIII'}(\tilde\gamma_+,\hat\gamma_-,\tilde\eta) &= \hat\gamma_- Q_1 \wedge Q_2 - \tilde\gamma_+ |\Lambda|^{-1/2} \big(Q_1 \wedge T_1 +  Q_2 \wedge T_2\big) - \frac{\tilde\eta}{2}\, |\Lambda|^{-1/2} J \wedge T_0\,.
\end{align}
The form of classes $r_{CII}$ and $r_{CV}$ was simplified here by anti-de Sitter-Carroll automorphisms $J \mapsto J - \hat\chi_+/\tilde\chi_-\, Q_2$, $T_1 \mapsto T_1 + \hat\chi_+/\tilde\chi_-\, T_0$ and $T_a \mapsto T_a + \hat\gamma/\tilde\varrho_- Q_a$, respectively, which also revealed that the parameters $\hat\chi_+$ and $\hat\gamma$ are actually redundant. The contraction limits of $r_{III''}$ and $r_{V''}$ are omitted, since both of them will turn out to be equal to $r_{CIIa}$ derived below (up to to a redefinition of parameters). 

Namely, as in the de Sitter case before, we should check whether acting with $\mathfrak{so}(2,2)$ algebra automorphisms on the anti-de Sitter $r$-matrices (\ref{eq:33.03a}-\ref{eq:36.03}) allows us to find any additional Carrollian (quantum) contraction limit. In this context, it needs to be mentioned that the 4D homogeneous space of the Lie group generated by the $\mathfrak{so}(2,2)$ algebra admits the flat metric $(-1,1,-1,1)$, while the 3D anti-de Sitter spacetime can be embedded in this space as a hypersurface orthogonal to a timelike or a spacelike axis. The automorphisms relevant for the considered extension of quantum contractions describe a change of that axis (cf. \cite{Kowalski:2020qs}) and a particular example of them is:
\begin{align}\label{eq:33.04gx}
{\cal J}_{1/2} \mapsto \pm|\Lambda|^{-1/2} {\cal P}_{1/2}\,, \quad {\cal J}_0 \mapsto -{\cal J}_0\,, \quad {\cal P}_{1/2} \mapsto \pm|\Lambda|^{1/2} {\cal J}_{1/2}\,, \quad {\cal P}_0 \mapsto -{\cal P}_0\,,
\end{align}
Examining all $r$-matrix classes, we find that it is sufficient to apply (\ref{eq:33.04gx}) to $r_{II}$ and $r_{IV}$, which then transform into
\begin{align}\label{eq:33.04g}
r_{II}^a(\chi_+,\chi_-,\varsigma) &= \frac{\chi_+}{2} \Big(\big({\cal J}_0 - {\cal J}_2\big) \wedge {\cal J}_1 - \tilde{\cal P}_1 \wedge \big(\tilde{\cal P}_0 - \tilde{\cal P}_2\big)\Big) \nonumber\\
&+ \frac{\chi_-}{2} \Big(\big({\cal J}_0 - {\cal J}_2\big) \wedge \tilde{\cal P}_1 - {\cal J}_1 \wedge \big(\tilde{\cal P}_0 - \tilde{\cal P}_2\big)\Big) - \frac{\varsigma}{2} \big({\cal J}_0 - {\cal J}_2\big) \wedge \big(\tilde{\cal P}_0 - \tilde{\cal P}_2\big)\,, \nonumber\\
r_{IV}^a(\gamma,\varsigma) &= \gamma\, \big({\cal J}_2 \wedge \tilde{\cal P}_0 - {\cal J}_0 \wedge \tilde{\cal P}_2 + {\cal J}_1 \wedge \tilde{\cal P}_1\big) - \frac{\varsigma}{2} \big({\cal J}_0 - {\cal J}_2\big) \wedge \big(\tilde{\cal P}_0 - \tilde{\cal P}_2\big)\,.
\end{align}
The Carrollian contraction limits of $r_{II}^a$ and $r_{IV}^a$ have the form
\begin{align}\label{eq:33.04ga}
r_{CIIa}(\hat\chi_+,\hat\chi_-,\hat\varsigma) &= \frac{\hat\chi_+}{2}\, Q_1 \wedge Q_2 + \frac{1}{2}\, |\Lambda|^{-1/2} \big(\hat\chi_- Q_1 - \hat\varsigma\, Q_2\big) \wedge T_0\,, \nonumber\\
r_{CIVa}(\tilde\gamma,\hat\gamma) &= -\tilde\gamma\, |\Lambda|^{-1/2} \big(J \wedge T_0 + Q_1 \wedge T_2 - Q_2 \wedge T_1\big) - \hat\gamma\, |\Lambda|^{-1/2} Q_2 \wedge T_0 \nonumber\\
&\cong -\tilde\gamma\, |\Lambda|^{-1/2} \big(J \wedge T_0 + Q_1 \wedge T_2 - Q_2 \wedge T_1\big) = r_{CIVa}(\tilde\gamma)\,.
\end{align}
In contrast to the automorphism (\ref{eq:32.04fx}) of de Sitter(-Carroll) algebra, the automorphism (\ref{eq:33.04gx}) does not correspond to an automorphism of anti-de Sitter-Carroll algebra and hence (\ref{eq:33.04ga}) provides genuinely new cases. As we already mentioned, $r_{CIIa}$ is actually the same $r$-matrix class as the previously omitted $r_{CIII''}$ and $r_{CV''}$. Explicitly, $r_{CIII''}(\hat\gamma,\hat{\bar\gamma},\hat\eta) = r_{CIIa}(\hat\gamma,\hat{\bar\gamma}/2,\hat\eta/2)$ and $r_{CV''}(\hat\gamma,\hat{\bar\chi},\hat\rho) = r_{CIIa}(\hat\gamma - \hat{\bar\chi}/2,-\hat{\bar\chi}/2,\hat\rho/2)$. The difference between $r_{CIV}$ and $r_{CIVa}$ is evident; the latter was obtained via the splitting and rescaling of $\varsigma$ analogous to $r_{CIV}$ of de Sitter-Carroll algebra (\ref{eq:32.01c}), as well as subsequently simplified by the automorphism $J \mapsto J - \hat\gamma/(2\tilde\gamma)\, Q_2$, $T_1 \mapsto T_1 + \hat\gamma/(2\tilde\gamma)\, T_0$ (the parameter $\hat\gamma$ turns out to be redundant). For completeness, let us also note that if $r_I$ is transformed by appropriate $\mathfrak{so}(2,2)$ automorphisms of the same kind as (\ref{eq:33.04gx}), its contraction limit will belong to the class $r_{CIIa}$ or $r_{CV}$. 

The inhomogeneous Yang-Baxter equations satisfied by the $r$-matrices (\ref{eq:35.03a},\ref{eq:33.04ga}) have the form
\begin{align}\label{eq:60.01ca}
[[r_{CIV},r_{CIV}]] &= [[r_{CIVa},r_{CIVa}]] = 2\tilde\gamma^2 \big(J \wedge Q_1 \wedge Q_2 + \Lambda^{-1} Q^a \wedge T_a \wedge T_0\big)\,, \nonumber\\
[[r_{CIII},r_{CIII}]] &= 2\tilde\gamma_-^2 \big(J \wedge Q_1 \wedge Q_2 + \Lambda^{-1} Q^a \wedge T_a \wedge T_0\big) - 4\tilde\gamma_- \hat\gamma_+ |\Lambda|^{-1/2} Q_1 \wedge Q_2 \wedge T_0\,, \nonumber\\
[[r_{CIII'},r_{CIII'}]] &= -2\tilde\gamma_+^2 \big(J \wedge Q_1 \wedge Q_2 + \Lambda^{-1} Q^a \wedge T_a \wedge T_0\big) + 4\tilde\gamma_+ \hat\gamma_- |\Lambda|^{-1/2} Q_1 \wedge Q_2 \wedge T_0\,,
\end{align}
while the equation is homogeneous in the remaining cases of (\ref{eq:33.03aa},\ref{eq:33.04ga}). The RHS-s of these equations consist of the same kind of terms as in (\ref{eq:60.02x}), in agreement with the expectation that the obtained quantum contraction limits are equivalent to the corresponding (sub)classes of $r$-matrices from our complete classification (\ref{eq:60.01x}). Indeed, a direct comparison shows that we can identify:
\begin{align}\label{eq:60.01y}
r_{CI}(\chi) &\cong r_{1'}(-\chi,\gamma = 0)\,, \nonumber\\
r_{CII}(\tilde\chi_-,\tilde\varsigma) &\cong r_{4'}(\chi = \tilde\chi_-/2,\varsigma = \tilde\varsigma/2,\theta = 0)\,, \nonumber\\
r_{CIIa}(\hat\chi_+,\hat\chi_-,\hat\varsigma) &\cong r_{8'}(\theta_{01} = \hat\varsigma/2,\theta_{12} = \hat\chi_+/2,\theta_{20} = -\hat\chi_-/2)\,, \nonumber\\
r_{CIII}(\hat\gamma_+,\tilde\gamma_-,\tilde\eta) &\cong r_{2'}(\gamma = \tilde\gamma_-,\eta = \tilde\eta/2,\theta_{20} = -\hat\gamma_+)\,, \nonumber\\
r_{CIII'}(\tilde\gamma_+,\hat\gamma_-,\tilde\eta) &\cong r_{3'}(\gamma = -\tilde\gamma_+,\eta = \tilde\eta/2,\theta_{12} = \hat\gamma_-)\,, \nonumber\\
r_{CIV}(\tilde\gamma,\tilde\varsigma) &\cong r_{6'}(\gamma = -\tilde\gamma,\varsigma = \tilde\varsigma/2,\theta = 0,\theta_{20} = 0)\,, \nonumber\\
r_{CIVa}(\tilde\gamma) &\cong r_{7'}(\gamma = \tilde\gamma)\,, \nonumber\\
r_{CV}(\hat\varrho_+,\tilde\varrho_-) &\cong r_{5'}(\chi = -\tilde\varrho_-/2,\theta_{01} = \hat\varrho_+/2,\theta_{12} = -\hat\varrho_+/2)\,.
\end{align}
In most of the cases, the identification is found after one acts with an automorphism describing an appropriate change of spatial axes. The conclusion is that (quantum) Carrollian contractions lead us to every class of the classification (\ref{eq:60.01x}). However, in contrast to the contractions relating deformations of de Sitter and de Sitter-Carroll algebras, they fail to recover some terms in four classes: $r_{1'}$, $r_{4'}$ and $r_{6'}$, as well as $r_{5'}$, for which $\theta_{12} = -\theta_{01}$. This is reflected in differences between the corresponding Yang-Baxter equations (\ref{eq:60.02x}) and (\ref{eq:60.01ca}). On the other hand, such a situation is similar to what we found in \cite{Kowalski:2020qs}, comparing the complete classification for Poincar\'{e} algebra with the $\Lambda \to 0$ quantum contraction limits of the complete classification for (anti-)de Sitter algebra (depicted in Fig.~\ref{fig:1}), where there was an even bigger number of the unrecovered terms. 

\begin{figure}[h]
\centering
\includegraphics{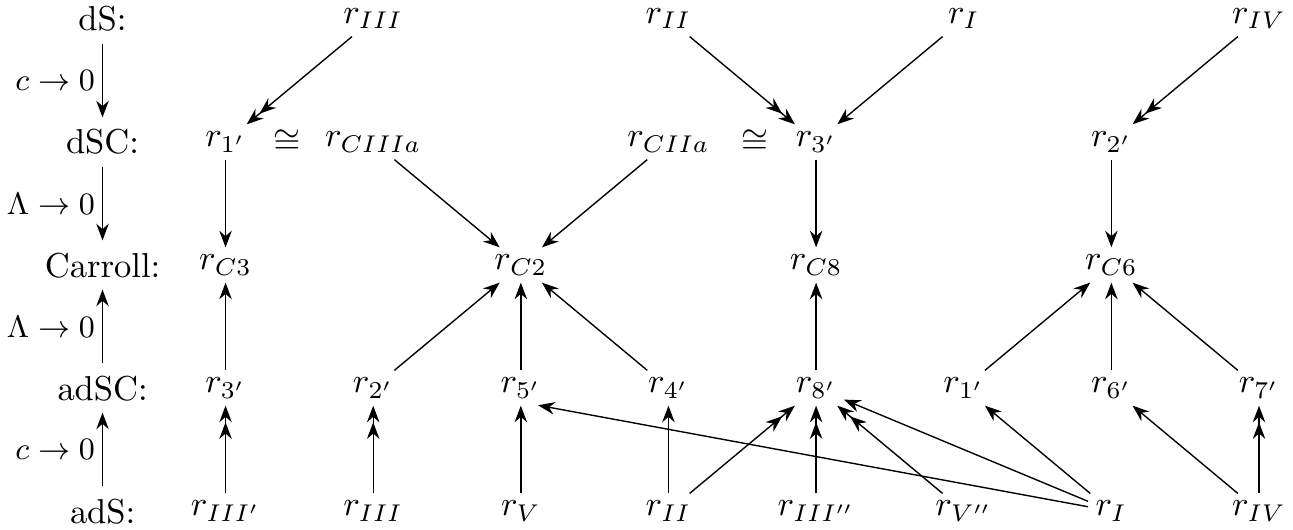}
\caption{\label{fig:2} Quantum ($c \to 0$ and $\Lambda \to 0$) contractions relating all $r$-matrix classes for de Sitter and anti-de Sitter algebras, all of those for dS-Carroll and adS-Carroll algebras, and those obtained for Carroll algebra; a two-headed arrow means that a given $c \to 0$ contraction recovers the full class (i.e., it is surjective).}
\end{figure}
 
Our discussion of classical $r$-matrices (characterizing possible bialgebras and hence quantum deformations) for the Carrollian kinematical algebras should be closed by considering the link connecting all of these algebras, which is the limit $\Lambda \to 0$. Namely, one may ask how the list of Carroll ($\Lambda = 0$) $r$-matrices (\ref{eq:60.01ba}) obtained from the classification of Poincar\'{e} $r$-matrices via (quantum) $c \to 0$ contractions compares to the results of (quantum) $\Lambda \to 0$ contractions applied to the classification of (anti-)de Sitter-Carroll $r$-matrices (\ref{eq:60.01x})/(\ref{eq:A0.02a}). Since the procedure follows the same patterns as for other contractions discussed in the current paper or in \cite{Kowalski:2020qs}, let us restrict to presenting the results on a diagram. We only need to mention that the contraction of $r_{6'}$ is performed with a splitting and rescaling of $\varsigma$: $\varsigma = (\varsigma - \gamma) + \gamma \equiv |\Lambda|\, \hat\gamma + |\Lambda|^{1/2} \tilde\gamma$. The diagram in Fig.~\ref{fig:2} depicts all $c \to 0$ contractions relating the (anti-)de Sitter and (anti-)de Sitter-Carroll $r$-matrices, as well as the most general $\Lambda \to 0$ contractions leading from the latter to the Carroll $r$-matrices. For clarity of the figure, we do not show that $r_{1'}(\gamma = 0)$ for (anti-)de Sitter-Carroll algebra contracts to $r_{C8}$, while $r_{6'}(\varsigma = 0)$ and $r_{8'}(\theta_{12} = 0)$ contract to $r_{C2}$. We also do not indicate that the $\Lambda \to 0$ contractions miss some terms proportional to the parameters $\theta_{\mu\nu}$.

\section{(A)dS-Galilei $r$-matrices from quantum contractions}\label{sec:5.0}

Last but not least, we will turn to deriving quantum contractions in the $c \to \infty$ limit of $r$-matrices for (anti-)de Sitter algebra, which will provide us with $r$-matrices for (anti-)de Sitter-Galilei algebra (\ref{eq:11e}). Based on the cases of de Sitter-Carroll and anti-de Sitter-Carroll algebras, it may be tentatively expected that such contractions allow us to recover the (unknown) complete classification for each of the considered algebras, possibly up to a few missing terms in some classes.

\subsection{The de Sitter case}

Performing Galilean quantum contractions of the de Sitter $r$-matrices (\ref{eq:32.01b}) in the same manner it was done for Poincar\'{e} algebra in (\ref{eq:60.01bc}), as well as dropping the (irrelevant) terms proportional to the antisymmetric split-Casimirs (\ref{eq:11cx}) and (\ref{eq:11ex}) (hence some $r$-matrices depend on fewer parameters than their de Sitter progenitors), we are left with the following independent classes:
\begin{align}\label{eq:32.01d}
r_{GI}(\tilde\chi) &= -\tilde\chi\, \big(Q_2 - \Lambda^{-1/2} T_2\big) \wedge \Lambda^{-1/2} T_0\,, \nonumber\\
r_{GII}(\tilde\chi,\hat\varsigma) &= \frac{\tilde\chi}{2} \Big(J \wedge \big(Q_1 - \Lambda^{-1/2} T_1\big) - \big(Q_2 - \Lambda^{-1/2} T_2\big) \wedge \Lambda^{-1/2} T_0\Big) \nonumber\\
&+ \frac{\hat\varsigma}{2} \big(Q_1 - \Lambda^{-1/2} T_1\big) \wedge \big(Q_2 - \Lambda^{-1/2} T_2\big)\,, \nonumber\\
r_{GIII}(\eta) &= 
- \frac{\eta}{2}\, \Lambda^{-1/2} J \wedge T_0\,. 
\end{align}
The class $r_{IV}(\gamma,\varsigma)$ reduced to $r_{GIV}(\hat\varsigma) = r_{GII}(\tilde\chi = 0,\hat\varsigma)$. 

Similarly to the case of Carrollian quantum contractions, we also need to consider if any additional contraction limits can be obtained after the $r$-matrices (\ref{eq:32.01b}) are first transformed by $\mathfrak{so}(3,1)$ automorphisms like the one given in (\ref{eq:32.04fx}). Since (as one can check) the latter is not inherited by de Sitter-Galilei algebra, this kind of automorphisms could lead to new Galilean contraction limits. Indeed, the transformed $r$-matrices (\ref{eq:32.04f}), as well as two additional ones,
\begin{align}\label{eq:32.04j}
r_I^a(\chi) &= \chi\, {\cal J}_1 \wedge \big(\tilde{\cal P}_0 - \tilde{\cal P}_2\big)\,, \nonumber\\
r_I^b(\chi) &= \chi\, \big({\cal J}_0 - {\cal J}_1\big) \wedge {\cal J}_2 
\end{align}
(the index $b$ denotes that we act with the inverse of (\ref{eq:32.04fx}), which corresponds to the choice of another spatial axis in the embedding 3+1-dimensional space), lead to dS-Galilei $r$-matrices inequivalent to any of the previous results (\ref{eq:32.01d}):
\begin{align}\label{eq:60.01bd}
r_{GIa}(\hat\chi) &= -\hat\chi\, \Lambda^{-1/2} Q_1 \wedge T_1\,, \nonumber\\
r_{GIb}(\hat\chi) &= -\hat\chi\, Q_1 \wedge Q_2 \cong -\hat\chi\, \Lambda^{-1} T_1 \wedge T_2\,, \nonumber\\
r_{GIVa}(\hat\gamma,\hat\varsigma) &= \hat\gamma\, \Lambda^{-1/2} Q_1 \wedge T_2 + \frac{\hat\varsigma}{2}\, \Lambda^{-1/2} Q_2 \wedge T_1\,, \nonumber\\
r_{GIIIa}(\tilde\gamma_-,\tilde\gamma_+,\hat\eta) &= \tilde\gamma_- \Lambda^{-1/2} \big(J \wedge T_1 + Q_2 \wedge T_0\big) - \tilde\gamma_+ \big(J \wedge Q_2 + \Lambda^{-1} T_0 \wedge T_1\big) \nonumber\\
&- \frac{\hat\eta}{2}\, \Lambda^{-1/2} Q_1 \wedge T_2\,.
\end{align}
We also wrote here an alternative form of $r_{GIb}$, given by the dS-Galilei algebra automorphism (\ref{eq:11ey}) with $\beta = 1$. Moreover, acting with the automorphism (\ref{eq:32.04fx}) (or its inverse), one can obtain a contraction limit of $r_{II}$ that belongs to the class $r_{GIVa}$. (These subtleties will be relevant for the diagram in Fig.~\ref{fig:3}). 

All of the $r$-matrices (\ref{eq:32.01d},\ref{eq:60.01bd}) satisfy the homogeneous Yang-Baxter equation, with the notable exception of
\begin{align}\label{eq:60.02cb}
[[r_{GIIIa},r_{GIIIa}]] &= -2 \big(\tilde\gamma_-^2 - \tilde\gamma_+^2\big) \Big(J \wedge Q_1 \wedge Q_2 - \Lambda^{-1} \big(J \wedge T_1 \wedge T_2 - Q^a \wedge T_a \wedge T_0\big)\Big) \nonumber\\
&+ 4\tilde\gamma_- \tilde\gamma_+ \Lambda^{-1/2} \big(J \wedge Q^a \wedge T_a - Q_1 \wedge Q_2 \wedge T_0 + \Lambda^{-1} T_0 \wedge T_1 \wedge T_2\big)\,.
\end{align}
We have already seen the same situation for Galilei algebra in (\ref{eq:60.02c}). In fact, $r_{GIIIa}$ is the contraction limit of $r_{III}^a \cong r_{III}$, which for $\gamma_- \neq 0$ or $\gamma_+ \neq 0$ describes the (twisted, generalized) de Sitter version of $\kappa$-deformation in a spatial direction. We will return to that in Sec.~\ref{sec:6.0}.

\subsection{The anti-de Sitter case}

Galilean quantum contractions of the anti-de Sitter $r$-matrices (\ref{eq:33.03a}-\ref{eq:36.03}) are performed analogously to the de Sitter ones (and again, some contraction limits will depend on fewer parameters than their anti-de Sitter progenitors due to dropping of the terms proportional to the antisymmetric split-Casimirs (\ref{eq:11cx}) and (\ref{eq:11ex})). Consequently, we obtain independent classes:
\begin{align}\label{eq:60.01be}
r_{GI}(\hat\chi) &= -\hat\chi\, |\Lambda|^{-1} T_1 \wedge T_2\,, \nonumber\\
r_{GIV}(\hat\gamma,\hat\varsigma) &= -\hat\gamma\, |\Lambda|^{-1/2} Q_1 \wedge T_2 - \frac{\hat\varsigma}{2}\, |\Lambda|^{-1/2} Q_2 \wedge T_1\,, \nonumber\\
r_{GIII'}(\eta) &= 
- \frac{\eta}{2}\, |\Lambda|^{-1/2} J \wedge T_0
\end{align}
and
\begin{align}\label{eq:60.01bf}
r_{GIII}(\tilde\gamma_-,\tilde\gamma_+,\hat\eta) &= \tilde\gamma_- \big(J \wedge Q_2 + \Lambda^{-1} T_0 \wedge T_1\big) - \tilde\gamma_+ |\Lambda|^{-1/2} \big(J \wedge T_1 + Q_2 \wedge T_0\big) \nonumber\\
&+ \frac{\hat\eta}{2}\, |\Lambda|^{-1/2} Q_1 \wedge T_2\,, \nonumber\\
r_{GV}(\tilde\gamma,\hat\varrho_+,\hat\varrho_-) &= \frac{\tilde\gamma}{2} \Big(J \wedge Q_2 + \Lambda^{-1} T_0 \wedge T_1 - |\Lambda|^{-1/2} \big(J \wedge T_1 + Q_2 \wedge T_0\big)\Big) \nonumber\\
&- \frac{1}{2} \big(\hat\varrho_+ Q_1 + \hat\varrho_- |\Lambda|^{-1/2} T_2\big) \wedge \big(Q_2 + |\Lambda|^{-1/2} T_1\big)\,, \nonumber\\
r_{GIII''}(\tilde{\bar\gamma},\tilde\eta) &= 
- \frac{\tilde{\bar\gamma}}{2} \Big(J \wedge Q_2 + \Lambda^{-1} T_0 \wedge T_1 + |\Lambda|^{-1/2} \big(J \wedge T_1 + Q_2 \wedge T_0\big)\Big) \nonumber\\
&- \frac{\tilde\eta}{2} \Big(J \wedge Q_1 + \Lambda^{-1} T_0 \wedge T_2 - |\Lambda|^{-1/2} \big(J \wedge T_2 + Q_1 \wedge T_0\big)\Big)\,. 
\end{align} 
The remaining $r$-matrices $r_{II}(\chi_+,\chi_-,\varsigma)$ and $r_{V''}(\gamma,\bar\chi,\rho)$ reduced to $r_{GII}(\hat\varsigma) = r_{GIV}(\hat\gamma = 0,\hat\varsigma)$ and $r_{GV''}(\tilde\rho) \cong r_{GIII''}(\tilde{\bar\gamma} = 0,\tilde\rho/2)$ (with the help of an automorphism rotating the spatial axes in the latter case). 

\begin{figure}[h]
\centering
\includegraphics{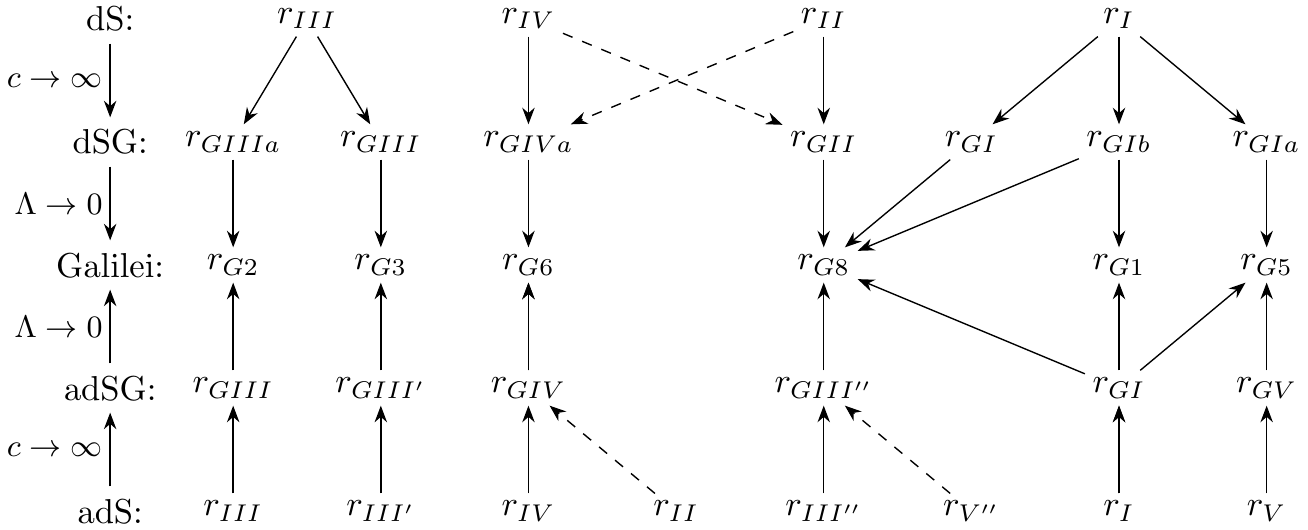}
\caption{\label{fig:3} Quantum ($c \to \infty$ and $\Lambda \to 0$) contractions relating all $r$-matrix classes for de Sitter and anti-de Sitter algebras, those obtained for dS-Galilei and adS-Galilei algebras, and those obtained for Galilei algebra; a dashed line means that a given $c \to \infty$ contraction leads to a subclass of a larger class.}
\end{figure}

In contrast to the Carrollian case, $\mathfrak{so}(2,2)$ algebra automorphisms that could allow us to obtain additional contraction limits (i.e., automorphisms describing a change of axis in the embedding 4D space, an example of which is (\ref{eq:33.04gx})) turn out to be inherited by anti-de Sitter-Galilei algebra and hence do not lead to any $r$-matrix not already contained in (\ref{eq:60.01be}-\ref{eq:60.01bf}). We can only note that if the contraction of $r_I$ is performed after transforming it by (\ref{eq:33.04gx}) or another map of this kind, e.g.:
\begin{align}\label{eq:33.04kx}
{\cal J}_{0/2} \mapsto \mp|\Lambda|^{-1/2} {\cal P}_{0/1}\,, \quad {\cal J}_1 \mapsto {\cal J}_2\,, \quad {\cal P}_{0/2} \mapsto \pm|\Lambda|^{1/2} {\cal J}_{0/1}\,, \quad {\cal P}_1 \mapsto -{\cal P}_2\,,
\end{align}
the result is an alternative form of $r_{GI}$, to wit
\begin{align}\label{eq:60.01bh}
r_{GI}(\hat\chi) \cong -\hat\chi\, Q_1 \wedge Q_2 \cong \hat\chi\, |\Lambda|^{-1/2} Q_1 \wedge T_1
\end{align}
(which will be relevant for the diagram in Fig.~\ref{fig:3}). 
 
The $r$-matrices (\ref{eq:60.01be}) satisfy the homogeneous Yang-Baxter equation, while for (\ref{eq:60.01bf}) we find
\begin{align}\label{eq:60.02cc}
[[r_{GIII},r_{GIII}]] &= 2 \big(\tilde\gamma_-^2 + \tilde\gamma_+^2\big) \Big(J \wedge Q_1 \wedge Q_2 - \Lambda^{-1} \big(J \wedge T_1 \wedge T_2 - Q^a \wedge T_a \wedge T_0\big)\Big) \nonumber\\
&+ 4\tilde\gamma_- \tilde\gamma_+ |\Lambda|^{-1/2} \big(J \wedge Q^a \wedge T_a - Q_1 \wedge Q_2 \wedge T_0 + \Lambda^{-1} T_0 \wedge T_1 \wedge T_2\big)\,, \nonumber\\
[[r_{GV},r_{GV}]] &= \tilde\gamma^2 \Big(J \wedge Q_1 \wedge Q_2 - \Lambda^{-1} \big(J \wedge T_1 \wedge T_2 - Q^a \wedge T_a \wedge T_0\big) \nonumber\\
&+ |\Lambda|^{-1/2} \big(J \wedge Q^a \wedge T_a - Q_1 \wedge Q_2 \wedge T_0\big) - |\Lambda|^{-3/2} T_0 \wedge T_1 \wedge T_2\Big)\,, \nonumber\\
[[r_{GIII''},r_{GIII''}]] &= \tilde{\bar\gamma}^2 \Big(J \wedge Q_1 \wedge Q_2 - \Lambda^{-1} \big(J \wedge T_1 \wedge T_2 - Q^a \wedge T_a \wedge T_0\big) \nonumber\\
&- |\Lambda|^{-1/2} \big(J \wedge Q^a \wedge T_a - Q_1 \wedge Q_2 \wedge T_0\big) + |\Lambda|^{-3/2} T_0 \wedge T_1 \wedge T_2\Big)\,.
\end{align}
Similarly as in the previous Subsection, we observe that the $r$-matrix describing the (twisted, generalized) $\kappa$-deformation in the time direction, i.e. $r_{GIII'}$, now satisfies the homogeneous equation, while the one describing the (twisted, generalized) $\kappa$-deformation in a spatial direction, i.e. $r_{GIII}$, still satisfies an inhomogeneous equation. There are also two other $r$-matrices being the solutions of inhomogeneous equations, $r_{GV}$ and $r_{GIII''}$, obtained from $r$-matrices that have no counterparts for de Sitter algebra. However, the inhomogenity comes from the terms of $r_{GV}$ and $r_{GIII''}$ that have the same form as the terms responsible for the inhomogenity of the equation solved by $r_{GIII}$.

Let us close the discussion of classical $r$-matrices (characterizing coboundary bialgebras) for the Galilean kinematical algebras in the same manner as in the Carrollian case, by considering the limit $\Lambda \to 0$. Namely, we compare the list of ($\Lambda = 0$) Galilei $r$-matrices (\ref{eq:60.01bc}) derived from the classification of Poincar\'{e} $r$-matrices via (quantum) $c \to \infty$ contractions to the results of (quantum) $\Lambda \to 0$ contractions of (anti-)de Sitter-Galilei $r$-matrices obtained in (\ref{eq:32.01d},\ref{eq:60.01bd}) and (\ref{eq:60.01be}-\ref{eq:60.01bf}). The diagram in Fig.~\ref{fig:3} depicts all $c \to \infty$ contractions leading from the (anti-)de Sitter to (anti-)de Sitter-Galilei $r$-matrices, as well as the most general $\Lambda \to 0$ contractions relating the latter with the Galilei $r$-matrices. As we previously explained, additional $c \to \infty$ contraction limits of $r_I$, $r_{II}$, $r_{III}$ and $r_{IV}$ for de Sitter algebra are obtained with the help of $\mathfrak{so}(3,1)$ automorphisms. Similarly, two equivalent forms of $r_{GIb}$ for dS-Galilei algebra (cf.~(\ref{eq:60.01bd})) and three equivalent forms of $r_{GI}$ for adS-Galilei algebra (cf.~(\ref{eq:60.01be}) and (\ref{eq:60.01bh})) lead to inequivalent $\Lambda \to 0$ contraction limits. We do not indicate in the figure that the $\Lambda \to 0$ contractions miss some terms proportional to the parameters $\theta_{\mu\nu}$ and that they can not recover $r_{G1}$ with $\hat\gamma \neq 0$. 

\section{Conclusions}\label{sec:6.0}

We studied classical $r$-matrices that characterize (coboundary) Hopf-algebraic deformations of the Carrollian and Galilean versions of the 2+1-dimensional Lorentzian kinematical algebras, i.e. Poincar\'{e} and (anti-)de Sitter algebras. In particular, the complete classification of such deformations for de Sitter-Carroll/anti-de Sitter-Carroll algebra, containing three/eight classes, is easily obtained by applying its isomorphism with Euclidean/Poincar\'{e} algebra to the corresponding complete classification from the literature. All deformations of the Carrollian algebras with $\Lambda \neq 0$ (up to a few terms missing in four $r$-matrix classes for anti-de Sitter-Carroll algebra, as demonstrated in (\ref{eq:60.01y})) can also be recovered via quantum $c \to 0$ contractions of the completely classified deformations of de Sitter and anti-de Sitter algebras. By analogy, it can be conjectured that seven/six classes of deformations of de Sitter-Galilei/anti-de Sitter-Galilei algebra derived by quantum $c \to \infty$ contractions of deformations of de Sitter/anti-de Sitter algebra recover (almost) all cases from the uknown complete classification. Meanwhile, both quantum $c \to 0$ contractions of Poincar\'{e} deformations and quantum $\Lambda \to 0$ contractions of (anti-)de Sitter-Carroll deformations lead to four $r$-matrix classes for Carroll algebra, up to some missing terms in some classes in the latter case. Both quantum $c \to \infty$ contractions of Poincar\'{e} deformations and quantum $\Lambda \to 0$ contractions of (anti-)de Sitter-Galilei deformations allow to similarly obtain six $r$-matrix classes for Galilei algebra, up to some missing terms in some classes in the latter case. However, it is possible that there are exist some additional coboundary deformations of Carroll or Galilei algebra, which can not be obtained by quantum contractions. 

The most physically interesting deformations of Poincar\'{e} algebra, which also have their counterparts for both de Sitter and anti-de Sitter algebras, are time-, light- and spacelike $\kappa$-deformations, as well as the one coming from the ``standard'' Drinfeld double structure (case 0 in \cite{Ballesteros:2019te}), the Lie bialgebra of which is often called the classical double, see e.g. \cite{Meusburger:2008qr}. Moreover, the corresponding $r$-matrices, apart from the lightlike case, stand out as the ones that survive as distinct classes under (almost) all quantum contractions relating deformations of the considered kinematical algebras, i.e. contractions in the limits $\Lambda \to 0$, $c \to 0$ and $c \to \infty$, as it may be traced with the help of Figs.~\ref{fig:1}-\ref{fig:3}. Table~\ref{tab:01} collects $r$-matrix (sub)classes corresponding to these special cases of deformations for each of the considered kinematical algebras (as it was mentioned in Subsec.~\ref{sec:2.2}, both $r_{III'}$ and $r_{III}$ -- and hence also $r_{GIII}$, contain two copies of a given $\kappa$-deformation, which can be transformed into each other using the respective automorphism of adS(G) algebra, (\ref{eq:33.04kx}) or (\ref{eq:33.04gx})). Strictly speaking, the scope of this paper was restricted to antisymmetric $r$-matrices, while $r$-matrices characterizing the classical double contain also the symmetric term, but the latter survives the Carrollian contractions and loses the products of $J$ and $T_0$ in the Galilean contractions, just like the antisymmetric ones. 

\begin{table}[h]
\centering
\begin{tabular}{|c|cc|cc|cc|}
\hline algebra & \multicolumn{2}{c|}{timelike $\kappa$-deformation} & \multicolumn{2}{c|}{spacelike $\kappa$-deformation} & \multicolumn{2}{c|}{classical double} \\
\hline\hline dSC & $r_{CIII}(\tilde\gamma_+) \cong r_{1'}(\gamma)$ & & $r_{CIIIa}(\hat\gamma_-) \cong r_{1'}(\theta_{12})$ & \!\!\!(T) & $r_{CIV}(\tilde\gamma) \cong r_{2'}(\gamma)$ & \\
\hline dS & $r_{III}(\gamma_+)$ & & $r_{III}^a(\gamma_-) \cong r_{III}(\gamma_-)$ & & $r_{IV}(2\gamma = \varsigma)$ & \\
\hline dSG & $0$ & \!\!\!(T) & $r_{GIIIa}(\tilde\gamma_-)$ & & $r_{GIVa}(2\hat\gamma = -\hat\varsigma)$ & \!\!\!(T) \\
\hline Carroll & $r_{C3}(\tilde\gamma)$ & & $r_{C2}(\hat\gamma)$ & \!\!\!(T) & $r_{C6}(\tilde\gamma)$ & \\
\hline Poincar\'{e} & $r_3(\gamma)$ & & $r_2(\gamma)$ & & $r_7(\gamma)$ & \\
\hline Galilei & $0$ & \!\!\!(T) & $r_{G2}(\tilde\gamma)$ & & $r_{G6}(\hat\gamma = \hat\varsigma)$ & \!\!\!(T) \\
\hline adSC & $r_{CIII'}(\tilde\gamma_+) \cong r_{3'}(\gamma)$ & & $r_{CIII}(\hat\gamma_+) \cong r_{2'}(\theta_{20})$ & \!\!\!(T) & $r_{CIVa}(\tilde\gamma) \cong r_{7'}(\gamma)$ & \\
\hline adS & $r_{III'}(\gamma_+) \cong r_{III'}(\gamma_-)$ & & $r_{III}(\gamma_+) \cong r_{III}(\gamma_-)$ & & $r_{IV}(2\gamma = -\varsigma)$ & \\
\hline adSG & $0$ & \!\!\!(T) & $r_{GIII}(\tilde\gamma_+) \cong r_{GIII}(\tilde\gamma_-)$ & & $r_{GIV}(2\hat\gamma = -\hat\varsigma)$ & \!\!\!(T) \\
\hline
\end{tabular}
\caption{$r$-matrices that characterize special cases of symmetry deformations, depending on a kinematical algebra (the omitted deformation parameters are equal to 0); (T) denotes that a given $r$-matrix is triangular, while otherwise it is quasitriangular.}\label{tab:01}
\end{table}

Based on that, we may observe that $\kappa$-deformation in the time direction is characterized by a quasitriangular $r$-matrix for all kinematical algebras apart from the Galilean ones, for which the  $r$-matrix is proportional to an antisymmetric split-Casimir, i.e. equivalent to 0 (and hence is trivially triangular). On the other hand, a study \cite{Ballesteros:2020ts} of its 3+1-dimensional version showed that the Galilean contraction of the corresponding Lie bialgebra can also be performed without any rescaling of the deformation parameter, leading to a Lie bialgebra with non-trivial cobrackets but which is not coboundary (i.e., does not have a $r$-matrix). Non-coboundary bialgebras exist for the Galilean kinematical algebras due to them being neither semisimple, nor inhomogeneous (pseudo-)orthogonal. They also deserve to be studied, but are less appealing from the theoretical perspective because there is no general prescription for their quantization. 

Meanwhile, $\kappa$-deformation in a spatial direction is characterized by a quasitriangular $r$-matrix for all kinematical algebras apart from the Carrollian ones, for which the $r$-matrix reduces to a (triangular) term proportional to $Q_2 \wedge T_0$. This apparent complementarity of the time- and spacelike $\kappa$-deformations has already been pointed out by us in Subsec.~\ref{sec:3.2}. The physical explanation could be that since the Carroll limit leads to a decoupling between points of space (the ultralocality mentioned in the Introduction), it makes the spacelike deformation somehow milder, while time becoming absolute in the Galilei limit neutralizes the (coboundary) timelike deformation. Even more significantly, light cones collapse in the Carroll limit and flatten out in the Galilei limit, which helps to understand why lightlike $\kappa$-deformation (described by $r_4(\chi)$ / $r_{II}^a(\chi)$ / $r_{II}^a(\chi_-)$ in the Poincar\'{e} / de Sitter / anti-de Sitter case) does not survive as a distinct deformation in either of these limits but, actually, converges with the spacelike deformation in the Carroll limit and the timelike deformation in the Galilei limit. However, as it was shown \cite{Ballesteros:2020ts} in 3+1 dimensions, the noncommutative geometry of spacetime associated with a deformation of the Carrollian or Galilean kinematical symmetries may still involve some mixing between time and spatial coordinates. 

Thirdly, a quasitriangular $r$-matrix associated with the classical double is preserved for all kinematical algebras apart from the Galilean ones, for which it reduces to a (triangular) $r$-matrix proportional to $Q_1 \wedge T_2 - Q_2 \wedge T_1$. Although this deformation does not distinguish any direction in spacetime, it turns out to be affected less by the Carrollian contractions. Taking into account that the Galilei limit had also a stronger neutralizing effect on the timelike $\kappa$-deformation than the Carroll limit on the spacelike $\kappa$-deformation, as well as comparing some other cases, allows us to conclude that deformations of the Galilean algebras are milder than deformations of the Carrollian algebras, which are in turn a bit milder than deformations of the Lorentzian algebras. 

Let us end the paper by briefly outlining possible applications of the obtained results. As we mentioned in Introduction, it has been shown that the Chern-Simons formulation of the theory of classical (2+1)d gravity generalizes to the gauge groups generated by non-Lorentzian kinematical algebras. At least for the algebras that we considered, the Poisson structure can still be described in terms of compatible $r$-matrices. A subtlety in the Galilean cases, as it has been shown already in \cite{Papageorgiou:2009as,Papageorgiou:2010ga}, is that the construction of Chern-Simons theory requires a so-called double extension of the kinematical algebra (interestingly, theories with this and further extensions are equivalent to certain Ho\v{r}ava-Lifshitz gravities \cite{Hartong:2016ny}). Consequently, our $r$-matrices would need to be supplemented by terms depending on the additional generators of the latter. The same obstacle does not concern Carroll \cite{Matulich:2019ln} and (due to the isomorphisms that we discussed before) a(dS)-Carroll algebras. The next step could be to analyze the Fock-Rosly compatibility of the corresponding $r$-matrices analogously to what has been done by us in \cite{Kowalski:2020qs}, followed by an analysis of kinematics determined by them. 

Meanwhile, not restricting ourselves to 2+1 dimensions, it is still uncertain if and how one can introduce Yang-Baxter deformations of a string (which is another area where $r$-matrices for the Lorentzian algebras are applied, see Introduction) in the non-Lorentzian settings. Some groundwork has been laid in the Galilean case \cite{Fontanella:2022cs}, while the Carrollian one is even less explored. The situation is complicated by the fact that strings are no longer described here by the well-understood sigma models. The other side of the coin is that one of proposals for the dual in the flat-space holography is a field theory with (conformal) Carroll symmetry. This is the so-called Carrollian holography, which contrasts with the celestial one \cite{Donnay:2023by}. In general, it remains to be seen whether quantum deformations could also help to illuminate the conformal side of the correspondence. 

An application in the technical sense would be to extend our study of quantum-deformed Carrollian/Galilean algebras to 3+1 (or more) dimensions, which seems straightforward, although with a limitation coming from the present lack of knowledge of the complete classification of $r$-matrices for higher-dimensional Lorentzian algebras. This avenue has already been opened -- we referred a few times here to the results obtained by Ballesteros et al. for the non-Lorentzian versions of (timelike) $\kappa$-Poincar\'{e} and $\kappa$-(anti-)de Sitter algebras in 3+1 dimensions. As one could expect, the overall effect of the Carrollian or Galilean contractions on quantum deformations does not depend on the number of dimensions, although the available deformations are a bit different. Another direction of generalization of our results, which we are currently pursuing, is to derive the non-Lorentzian versions of quantum deformations of BMS and $\Lambda$-BMS algebras that have been constructed by Borowiec et al. (see Introduction). It is worth to mention that while all $r$-matrices playing a role in (2+1)d gravity belong to the quasitriangular ones, deformations of the asymptotic symmetry algebras are only obtained from the triangular ones.

\section*{Acknowledgments}
The author expresses gratitude for fruitful discussions with A.~Borowiec and J.~Kowalski-Glikman, as well as valuable communications from S.~J.~van Tongeren. This work is supported by the National Science Center, project no. UMO-2022/45/B/ST2/01067.

\appendix
\section{Embedding of (2+1)d Carroll in (3+1)d Poincar\'{e} algebra}\label{sec:A}

As we mentioned in Introduction, Carroll algebra can also be defined as the algebra of symmetries of a null hypersurface in Minkowski spacetime one dimension higher. Therefore, it should be directly related to symmetries of the latter. This relation is indeed described in e.g. \cite{Figueroa:2023ly} but here we would like to show it in the 2+1-dimensional case in terms of the explicit formulae. 

In 3+1 dimensions, Poincar\'{e} algebra $\mathfrak{iso}(3,1) = \mathfrak{so}(3,1) \vartriangleright\!\!< \mathbbm{R}^{3,1}$ expressed in the basis analogous to (\ref{eq:10}) has the brackets
\begin{align}\label{eq:08}
[J_i,J_j] & = \epsilon_{ij}^{\ \ k} J_k\,, & [J_i,K_j] & = \epsilon_{ij}^{\ \ k} K_k\,, & [K_i,K_j] & = -\epsilon_{ij}^{\ \ k} J_k\,, \nonumber\\ 
[J_i,P_j] & = \epsilon_{ij}^{\ \ k} P_k\,, & [J_i,P_0] & = 0\,, & [P_i,P_j] & = 0\,, \nonumber\\ 
[K_i,P_j] & = \delta_{ij} P_0\,, & [K_i,P_0] & = P_i\,, & [P_0,P_i] & = 0
\end{align}
for the generators of rotations (i.e. elliptic Lorentz transformations) $J_i$, boosts (i.e. hyperbolic Lorentz transformations) $K_i$ and space-time translations $P_i,P_0$; $i = 1,2,3$ and indices are raised with Euclidean metric. Let us also define three pairs of generators of null rotations (i.e. parabolic Lorentz transformations),
\begin{align}\label{eq:02}
L^{(1)}_{2/3} = \pm J_{2/3} - K_{3/2}\,, \qquad L^{(2)}_{3/1} = \pm J_{3/1} - K_{1/3}\,, \qquad L^{(3)}_{1/2} = \pm J_{1/2} - K_{2/1}\,,
\end{align}
where the upper index indicates the null direction preserved by the corresponding transformations, $u^i = (t - x^i)/\sqrt{2}$, while the lower index -- the preserved spatial direction. An equivalent choice would be to consider the following three pairs of generators:
\begin{align}\label{eq:03}
L^{(1')}_{2/3} = \pm J_{2/3} + K_{3/2}\,, \qquad L^{(2')}_{3/1} = \pm J_{3/1} + K_{1/3}\,, \qquad L^{(3')}_{1/2} = \pm J_{1/2} + K_{2/1}\,,
\end{align}
which preserve the respective null directions $v^i = (t + x^i)/\sqrt{2}$. 

The 2+1-dimensional Carroll algebra (\ref{eq:11}) can be embedded as a subalgebra of (\ref{eq:08}) if we choose one pair of the null-rotation generators (\ref{eq:02}) and identify
\begin{align}\label{eq:04}
J := -J_i\,, \qquad Q_{1/2} := -L^{(i)}_{j/k}\,, \qquad T_0 := P_0 - P_i\,, \qquad T_{1/2} := P_{k/j}\,,
\end{align}
or choose one pair of the null-rotation generators (\ref{eq:03}) and identify
\begin{align}\label{eq:05}
J := -J_i\,, \qquad Q_{1/2} := -L^{(i')}_{j/k}\,, \qquad T_0 := P_0 + P_i\,, \qquad T_{1/2} := P_{k/j}\,,
\end{align}
where $i,j,k$ are given by an even permutation of $\{1,2,3\}$. For example, an embedding may be given by $-J_1$, $-L^{(1)}_2$, $-L^{(1)}_3$, $P_0 - P_1$, $P_3$ and $P_2$, respectively. 

Is there any significance of this in our context? Since the classification of $r$-matrices for the 3+1-dimensional Poincar\'{e} algebra is mostly known \cite{Zakrzewski:1997pp}, we could try to extract coboundary deformations of the 2+1-dimensional Carroll algebra from the former (possibly obtaining additional results with respect to quantum contractions in (\ref{eq:60.01ba})). It would involve choosing one of the embeddings (\ref{eq:04}-\ref{eq:05}) and finding such representatives of the $\mathfrak{iso}(3,1)$ $r$-matrix classes that depend only on the generators in the image of the embedding. Let us note that not all automorphisms of the considered Carroll algebra extend to the full $\mathfrak{iso}(3,1)$, hence some $r$-matrices obtained in this way may turn out to belong to the same equivalence class. Moreover, the RHS-s of Yang-Baxter equations (\ref{eq:60.02b}) do not map onto the invariants of $\mathfrak{iso}(3,1)$, which makes it more tricky to recover quasitriangular $r$-matrices. Such a study might be the subject of future work.


\begin{thebibliography}{99}

\bibitem{Bacry:1968ps}
H.~Bacry and J.~L\'{e}vy-Leblond, 
{\it Possible kinematics}, 
J.\ Math.\ Phys.\ {\bf 9}, 1605 (1968).

\bibitem{Bacry:1986cy}
H. Bacry and J. Nuyts, 
{\it Classification of ten-dimensional kinematical groups with space isotropy}, \linebreak
J.\ Math.\ Phys.\ {\bf 27}, 2455 (1986).


\bibitem{Figueroa:2018hy}
J.~Figueroa-O'Farrill, 
{\it Higher-dimensional kinematical Lie algebras via deformation theory}, \linebreak
J.\ Math.\ Phys.\ {\bf 59}, 061702 (2018) [arXiv:1711.07363 [hep-th]].

\bibitem{Andrzejewski:2018ks}
T.~Andrzejewski and J.~Figueroa-O'Farrill, 
{\it Kinematical lie algebras in 2 + 1 dimensions}, \linebreak
J.\ Math.\ Phys.\ {\bf 59}, 061703 (2018) [arXiv:1802.04048 [hep-th]].

\bibitem{Figueroa:2019ss}
J.~Figueroa-O'Farrill and S.~Prohazka, 
{\it Spatially isotropic homogeneous spacetimes}, \linebreak
J.\ High Energy Phys.\ {\bf 01}, 229 (2019) [arXiv:1809.01224 [hep-th]].

\bibitem{Bergshoeff:2023ar}
E.~Bergshoeff, J.~Figueroa-O'Farrill and J.~Gomis, 
{\it A non-lorentzian primer}, \linebreak
SciPost Phys.\ Lect.\ Notes {\bf 69}, 1 (2023) [arXiv:2206.12177 [hep-th]].

\bibitem{Figueroa:2022ns}
J.~Figueroa-O'Farrill, 
{\it Non-lorentzian spacetimes}, 
Differ.\ Geom.\ Appl.\ {\bf 82}, 101894 (2022) [arXiv:2204.13609 [math.DG]].

\bibitem{Perez:2021ay}
A.~P\'{e}rez, 
{\it Asymptotic symmetries in Carrollian theories of gravity}, 
J.\ High Energy Phys.\ {\bf 12}, 173 (2021) [arXiv:2110.15834 [hep-th]].

\bibitem{Fuentealba:2022as}
O.~Fuentealba, M.~Henneaux, P.~Salgado-Rebolledo and J.~Salzer, 
{\it Asymptotic structure of Carrollian limits of Einstein-Yang-Mills theory in four spacetime dimensions}, 
Phys.\ Rev.\ D {\bf 106}, 104047 (2022) [arXiv:2207.11359 [hep-th]].

\bibitem{Parsa:2019os}
A.~Farahmand Parsa, H.~R.~Safari and M.~M.~Sheikh-Jabbari, 
{\it On rigidity of 3d asymptotic symmetry algebras}, 
J.\ High Energy Phys.\ {\bf 03}, 143 (2019) [arXiv:1809.08209 [hep-th]].

\bibitem{Safari:2019os}
H.~R.~Safari and M.~M.~Sheikh-Jabbari, 
{\it BMS$_4$ algebra, its stability and deformations}, \linebreak
J.\ High Energy Phys.\ {\bf 04}, 68 (2019) [arXiv:1902.03260 [hep-th]].

\bibitem{Bergshoeff:2014ds}
E.~Bergshoeff, J.~Gomis and G.~Longhi, 
{\it Dynamics of Carroll particles},
Class.\ Quant.\ Grav.\ {\bf 31}, 205009 (2014) [arXiv:1405.2264 [hep-th]].

\bibitem{Cardona:2016ds}
B.~Cardona, J.~Gomis and J.~M.~Pons, 
{\it Dynamics of Carroll strings},
J.\ High Energy Phys.\ {\bf 07}, 50 (2016) [arXiv:1605.05483 [hep-th]].

\bibitem{Marsot:2022pn}
L.~Marsot, 
{\it Planar Carrollean dynamics, and the Carroll quantum equation},
J.\ Geom.\ Phys.\ {\bf 179}, 104574 (2022) [arXiv:2110.08489 [math-ph]].

\bibitem{Marsot:2023hs}
L.~Marsot, P.~M.~Zhang, M.~Chernodub and P.~A.~Horv\'{a}thy, 
{\it Hall motions in Carroll dynamics},
Phys.\ Rept.\ {\bf 1028}, 1 (2023) [arXiv:2212.02360 [hep-th]].




\bibitem{Duval:2014ce}
C.~Duval, G.~W.~Gibbons, P.~A.~Horv\'{a}thy and P.~M.~Zhang, 
{\it Carroll versus Newton and Galilei: two dual non-Einsteinian concepts of time},
Class.\ Quant.\ Grav.\ {\bf 31}, 085016 (2014) [arXiv:1402.0657 [gr-qc]].

\bibitem{Figueroa:2023ly}
J.~Figueroa-O'Farrill, 
{\it Lie algebraic Carroll/Galilei duality}, 
J.\ Math.\ Phys.\ {\bf 64}, 013503 (2023) [arXiv:2210.13924 [math.DG]].

\bibitem{Bergshoeff:2017cy}
E.~Bergshoeff, J.~Gomis, B.~Rollier, J.~Rosseel and T.~ter Veldhuis, 
{\it Carroll versus Galilei gravity}, \linebreak
J.\ High Energy Phys.\ {\bf 03}, 165 (2017) [arXiv:1701.06156 [hep-th]].

\bibitem{Henneaux:1979gs}
M.~Henneaux, 
{\it Geometry of Zero Signature Space-times}, 
Bull.\ Soc.\ Math.\ Belg.\ {\bf 31}, 47 (1979).

\bibitem{Henneaux:2021cs}
M.~Henneaux and P.~Salgado-Rebolled\'{o}, 
{\it Carroll contractions of Lorentz-invariant theories}, \linebreak
J.\ High Energy Phys.\ {\bf 11}, 180 (2021) [arXiv:2109.06708 [hep-th]].

\bibitem{Hansen:2022cy}
D.~Hansen, N.~A.~Obers, G.~Oling and B.~T.~S\o gaard, 
{\it Carroll Expansion of General Relativity}, \linebreak
SciPost Phys.\ {\bf 13}, 055 (2022) [arXiv:2112.12684 [hep-th]].


\bibitem{Isham:1976sy}
C.~J.~Isham, 
{\it Some quantum field theory aspects of the superspace quantization of general relativity}, 
Proc.\ Roy.\ Soc.\ Lond.\ A {\bf 351}, 209 (1976).


\bibitem{Belinsky:1970oy}
V.~A.~Belinsky, I.~M.~Khalatnikov and E.~M.~Lifshitz,
{\it Oscillatory approach to a singular point in the relativistic cosmology}, 
Adv.\ Phys.\ {\bf 19}, 525 (1970).


\bibitem{Andersson:2005as}
L.~Andersson, H.~van Elst, W.~C.~Lim and C.~Uggla,
{\it Asymptotic silence of generic cosmological singularities}, 
Phys.\ Rev.\ Lett.\ {\bf 94}, 051101 (2005) [gr-qc/0402051].

\bibitem{Mielczarek:2013ay}
J.~Mielczarek, 
{\it Asymptotic silence in loop quantum cosmology}, 
AIP Conf.\ Proc.\ {\bf 1514}, 81 (2013) [arXiv:1212.3527 [gr-qc]].

\bibitem{Mielczarek:2017se}
J.~Mielczarek and T.~Trze\'{s}niewski, 
{\it Spectral dimension with deformed spacetime signature}, \linebreak
Phys.\ Rev.\ D {\bf 96}, 024012 (2017) [arXiv:1612.03894 [hep-th]].

\bibitem{Morand:2020es}
K.~Morand, 
{\it Embedding Galilean and Carrollian geometries. I. Gravitational waves}, \linebreak
J.\ Math.\ Phys.\ {\bf 61}, 082502 (2020) [arXiv:1811.12681 [hep-th]].

\bibitem{Duval:2017cs}
C.~Duval, G.~W.~Gibbons, P.~A.~Horv\'{a}thy and P.~M.~Zhang, 
{\it Carroll symmetry of plane gravitational waves}, 
Class.\ Quant.\ Grav.\ {\bf 34}, 175003 (2017) [arXiv:1702.08284 [gr-qc]].

\bibitem{Donnay:2019cn}
L.\ Donnay and C.\ Marteau, 
{\it Carrollian physics at the black hole horizon}, 
Class.\ Quant.\ Grav.\ {\bf 36}, 165002 (2019) [arXiv:1903.09654 [hep-th]].

\bibitem{Duval:2014cs}
C.~Duval, G.~W.~Gibbons and P.~A.~Horv\'{a}thy, 
{\it Conformal Carroll groups and BMS}, \linebreak
Class.\ Quant.\ Grav.\ {\bf 31}, 092001 (2014) [arXiv:1402.5894 [gr-qc]].

\bibitem{Ciambelli:2019cs}
L.~Ciambelli, R.~G.~Leigh, C.~Marteau, and P.~M.~Petropoulos, 
{\it Carroll structures, null geometry, and conformal isometries}, 
Phys.\ Rev.\ D {\bf 100}, 046010 (2019) [arXiv:1905.02221 [hep-th]].

\bibitem{Lukierski:1991qa}
J.~Lukierski, H.~Ruegg, A.~Nowicki and V.~N.~Tolstoy, 
{\it q-deformation of Poincar\'{e} algebra}, \linebreak
Phys.\ Lett.\ B {\bf 264}, 331 (1991).

\bibitem{Lukierski:1992ny}
J.~Lukierski, A.~Nowicki and H.~Ruegg, 
{\it New quantum Poincar\'{e} algebra and $\kappa$-deformed field theory}, 
Phys.\ Lett.\ B {\bf 293}, 344 (1992).


\bibitem{Maslanka:1993tp}
P.~Ma\'{s}lanka, 
{\it The $n$-dimensional $\kappa$-Poincar\'{e} algebra and group}, 
J.\ Phys.\ A: Math.\ Gen.\ {\bf 26}, L1251 (1993).


\bibitem{Majid:1994by}
S.~Majid and H.~Ruegg, 
{\it Bicrossproduct structure of $\kappa$-Poincar\'e group and noncommutative geometry}, 
Phys.\ Lett.\ B {\bf 334}, 348 (1994) [hep-th/9405107].

\bibitem{Ballesteros:2017ts}
A.~Ballesteros, F.~J.~Herranz, F.~Musso and P.~Naranjo, 
{\it The $\kappa$-(A)dS quantum algebra in (3+1) dimensions}, 
Phys.\ Lett.\ B {\bf 766}, 205 (2017) [arXiv:1612.03169 [hep-th]].

\bibitem{Ballesteros:2019tf}
A.~Ballesteros, I.~Gutierrez-Sagredo and F.~J.~Herranz, 
{\it The $\kappa$-(A)dS noncommutative spacetime}, 
Phys.\ Lett.\ B {\bf 796}, 93 (2019) [arXiv:1905.12358 [math-ph]].

\bibitem{Ballesteros:1994qh}
A.~Ballesteros, F.~J.~Herranz, M.~A.~del Olmo and M.~Santander, 
{\it Quantum (2+1) kinematical algebras: a global approach}, 
J.\ Phys.\ A: Math.\ Gen.\ {\bf 27}, 1283 (1994).

\bibitem{Ballesteros:2017ns}
A.~Ballesteros, N.~R.~Bruno and F.~J.~Herranz, 
{\it Non-commutative relativistic spacetimes and worldlines from 2+1 quantum (anti-)de Sitter groups}, 
Adv.\ High Energy Phys.\ {\bf 2017}, 7876942 (2017) [hep-th/0401244].

\bibitem{Borowiec:2019ky}
A.~Borowiec, L.~Brocki, J.~Kowalski-Glikman and J.~Unger, 
{\it $\kappa$-deformed BMS symmetry}, \linebreak
Phys.\ Lett.\ B {\bf 790}, 415 (2019) [arXiv:1811.05360 [hep-th]].

\bibitem{Borowiec:2021bs}
A.~Borowiec, L.~Brocki, J.~Kowalski-Glikman and J.~Unger, 
{\it BMS algebras in 4 and 3 dimensions, their quantum deformations and duals}, 
J.\ High Energy Phys.\ {\bf 02}, 84 (2021) [arXiv:2010.10224 [hep-th]].

\bibitem{Borowiec:2021ds}
A.~Borowiec, J.~Kowalski-Glikman and J.~Unger, 
{\it 3-dimensional $\Lambda$-BMS symmetry and its deformations}, 
J.\ High Energy Phys.\ {\bf 11}, 103 (2021) [arXiv:2106.12874 [hep-th]].

\bibitem{Stachura:1998ps} P.~Stachura, 
{\it Poisson-Lie structures on Poincar\'{e} and Euclidean groups in three dimensions}, \linebreak
J.\ Phys.\ A: Math.\ Gen.\ {\bf 31}, 4555 (1998).

\bibitem{Zakrzewski:1994pp}
S.~Zakrzewski, 
{\it Poisson structures on the Lorentz group}, 
Lett.\ Math.\ Phys.\ {\bf 32}, 11 (1994).

\bibitem{Borowiec:2016qg}
A.~Borowiec, J.~Lukierski and V.~N.~Tolstoy, 
{\it Quantum deformations of $D = 4$ Euclidean, Lorentz, Kleinian and quaternionic $\mathfrak{o}^\star(4)$ symmetries in unified $\mathfrak{o}(4;\mathbbm{C})$ setting}, 
Phys.\ Lett.\ B {\bf 754}, 176 (2016) [arXiv:1511.03653 [hep-th]].

\bibitem{Borowiec:2017ag}
A.~Borowiec, J.~Lukierski and V.~N.~Tolstoy, 
{\it Addendum to ``Quantum deformations of $D = 4$ Euclidean, Lorentz, Kleinian and quaternionic $\mathfrak{o}^\star(4)$ symmetries in unified $\mathfrak{o}(4;\mathbbm{C})$ setting''}, \linebreak
Phys.\ Lett.\ B {\bf 770}, 426 (2017) [arXiv:1704.06852 [hep-th]].

\bibitem{Borowiec:2017bs}
A.~Borowiec, J.~Lukierski and V.~N.~Tolstoy, 
{\it Basic quantizations of $D = 4$ Euclidean, Lorentz, Kleinian and quaternionic $\mathfrak{o}^\star(4)$ symmetries}, 
J.\ High Energy Phys.\ {\bf 11}, 187 (2017) [arXiv:1708.09848 [hep-th]].

\bibitem{Kowalski:2020qs}
J.~Kowalski-Glikman, J.~Lukierski and T.~Trze\'{s}niewski, 
{\it Quantum D = 3 Euclidean and Poincar\'{e} symmetries from contraction limits}, 
J.\ High Energy Phys.\ {\bf 09}, 096 (2020), [arXiv:1911.09538 [hep-th]].

\bibitem{Zakrzewski:1997pp}
S.~Zakrzewski, 
{\it Poisson Structures on the Poincar\'{e} Group}, 
Comm.\ Math.\ Phys.\ {\bf 185}, 285 (1997) [q-alg/9602001].

\bibitem{Ballesteros:2022ns}
A.~Ballesteros, I.~Gutierrez-Sagredo and F.~J.~Herranz, 
{\it Noncommutative (A)dS and Minkowski spacetimes from quantum Lorentz subgroups}, 
Class.\ Quant.\ Grav.\ {\bf 39}, 015018 (2022) [arXiv:2108.02683 [math-ph]].

\bibitem{Oriti:2009ay}
D.~Oriti (ed.), 
``Approaches to Quantum Gravity,'' 
Cambridge Univ.\ Press 2009.

\bibitem{Addazi:2022qw}
A.~Addazi et al., 
{\it Quantum gravity phenomenology at the dawn of the multi-messenger era -- A review}, 
Prog.\ Part.\ Nucl.\ Phys.\ {\bf 125}, 103948 (2022) [arXiv:2111.05659 [hep-ph]].

\bibitem{Pachol:2016qg}
A.~Pacho\l\ and S.~J.~van Tongeren, 
{\it Quantum deformations of the
flat space superstring}, 
Phys.\ Rev.\ D {\bf 93}, 026008 (2016) [arXiv:1510.02389 [hep-th]].

\bibitem{Osten:2017as}
D.~Osten and S.~J.~van Tongeren, 
{\it Abelian Yang-Baxter deformations and TsT transformations}, \linebreak
Nucl.\ Phys.\ B {\bf 915}, 184 (2017) [arXiv:1608.08504 [hep-th]].

\bibitem{Idiab:2022yg}
K.~Idiab and S.~J.~van Tongeren, 
{\it Yang-Baxter deformations of the flat space string}, 
Phys.\ Lett.\ B {\bf 835}, 137499 (2022) [arXiv:2205.13050 [hep-th]].

\bibitem{Hoare:2022is}
B.~Hoare, 
{\it Integrable deformations of sigma models}, 
J.\ Phys.\ A: Math.\ Theor.\ {\bf 55}, 093001 (2022) [arXiv:2109.14284 [hep-th]].

\bibitem{Achucarro:1986as}
A.~Ach\'{u}carro and P.~K.~Townsend, 
{\it A Chern-Simons action for three-dimensional anti-de Sitter supergravity theories},
Phys.\ Lett.\ B {\bf 180}, 89 (1986).

\bibitem{Witten:1988dm}
E.~Witten, 
{\it 2+1 dimensional gravity as an exactly soluble system},
Nucl.\ Phys.\ B {\bf 311}, 46 (1988).

\bibitem{Fock:1999px}
V.~V.~Fock and A.~A.~Rosly, 
{\it Poisson structure on moduli of flat connections on Riemann surfaces and $r$-matrix}, 
Am.\ Math.\ Soc.\ Transl.\ {\bf 191}, 67 (1999) [math/9802054].

\bibitem{Meusburger:2008qr}
C.~Meusburger and B.~J.~Schroers, 
{\it Quaternionic and Poisson-Lie structures in three-dimensional gravity: The cosmological constant as deformation parameter}, 
J.\ Math.\ Phys.\ {\bf 49}, 083510 (2008) [arXiv:0708.1507 [gr-qc]].

\bibitem{Meusburger:2009gy}
C.~Meusburger and B.~J.~Schroers, 
{\it Generalised Chern-Simons actions for 3d gravity and $\kappa$-Poincar\'{e} symmetry}, 
Nucl.\ Phys.\ B {\bf 806}, 462 (2009) [arXiv:0805.3318 [gr-qc]].

\bibitem{Schroers:2011qh}
B.~J.~Schroers, 
{\it Quantum gravity and non-commutative spacetimes in three dimensions: a unified approach}, 
Acta Phys.\ Polon.\ B Proc.\ Suppl.\ {\bf 4}, 379 (2011) [arXiv:1105.3945 [gr-qc]].

\bibitem{Osei:2018cy}
P.~K.~Osei and B.~J.~Schroers, 
{\it Classical r-matrices for the generalised Chern-Simons formulation of 3d gravity}, 
Class.\ Quantum Grav.\ {\bf 35}, 075006 (2018) [arXiv:1708.07650 [hep-th]].

\bibitem{Osei:2012oy}
P.~K.~Osei and B.~J.~Schroers, 
{\it On the semiduals of local isometry groups in three-dimensional gravity}, 
J.\ Math.\ Phys.\ {\bf 53} 073510, (2012) [arXiv:1109.4086 [gr-qc]].

\bibitem{Rosati:2017kt}
G.~Rosati, 
{\it $\kappa$-de Sitter and $\kappa$-Poincar\'{e} symmetries emerging from Chern-Simons (2+1)D gravity with cosmological constant}, 
Phys.\ Rev.\ D {\bf 96}, 066027 (2017) [arXiv:1706.02868 [hep-th]].

\bibitem{Ballesteros:2013dy}
A.~Ballesteros, F.~J.~Herranz and C.~Meusburger, 
{\it Drinfeld doubles for (2+1)-gravity}, \linebreak
Class.\ Quantum Grav.\ {\bf 30}, 155012 (2013) [arXiv:1303.3080 [math-ph]].

\bibitem{Ballesteros:2019te}
A.~Ballesteros, I.~Gutierrez-Sagredo and F.~J.~Herranz, 
{\it The Poincar\'{e} group as a Drinfeld double}, 
Class.\ Quantum Grav.\ {\bf 36}, 025003 (2019) [arXiv:1809.09207 [math-ph]].

\bibitem{Matulich:2019ln}
J.~Matulich, S.~Prohazka and J.~Salzer, 
{\it Limits of three-dimensional gravity and metric kinematical Lie algebras in any dimension}, 
J.\ High Energy Phys.\ {\bf 07}, 118 (2019) [arXiv:1903.09165 [hep-th]].

\bibitem{Kowalski:2014dy}
J.~Kowalski-Glikman and T.~Trze\'{s}niewski, 
{\it Deformed Carroll particle from 2+1 gravity}, \linebreak
Phys.\ Lett.\ B {\bf 737}, 267 (2014) [arXiv:1408.0154 [hep-th]].

\bibitem{Trzesniewski:2018ey}
T.~Trze\'{s}niewski, 
{\it Effective Chern-Simons actions of particles coupled to 3D gravity}, \linebreak
Nucl.\ Phys.\ B {\bf 928}, 448 (2018), [arXiv:1706.01375 [hep-th]].

\bibitem{Trzesniewski:2023gs}
T.~Trze\'{s}niewski, 
{\it 3D Gravity, Point Particles, and Deformed Symmetries}, \linebreak
Acta Phys.\ Polon.\ B Proc.\ Suppl.\ {\bf 16}, 6-A19 (2023) [arXiv:2212.14031 [hep-th]].

\bibitem{Giller:1992ma}
S.~Giller, P.~Kosi\'{n}ski, M.~Majewski, P.~Ma\'{s}lanka and J.~Kunz, 
{\it More about the q-deformed Poincar\'{e} algebra}, 
Phys.\ Lett.\ B {\bf 286}, 57 (1992).

\bibitem{Daszkiewicz:2008cs}
M.~Daszkiewicz, 
{\it Canonical and Lie-algebraic twist deformations of $\kappa$-Poincar\'{e} and contractions to $\kappa$-Galilei algebras}, 
Int.\ J.\ Mod.\ Phys.\ A {\bf 23}, 4387 (2008) [arXiv:0802.1974 [math-ph]].

\bibitem{Daszkiewicz:2019cs}
M.~Daszkiewicz, 
{\it Canonical and Lie-algebraic twist deformations of Carroll, para-Galilei and Static Hopf algebras}, 
Mod.\ Phys.\ Lett.\ A {\bf 34}, 1950181 (2019).

\bibitem{Ballesteros:2020ts}
A.~Ballesteros, G.~Gubitosi, I.~Gutierrez-Sagredo and F.~J.~Herranz, 
{\it The $\kappa$-Newtonian and $\kappa$-Carrollian algebras and their noncommutative spacetimes}, 
Phys.\ Lett.\ B {\bf 805}, 135461 (2020) [arXiv:2003.03921 [hep-th]].

\bibitem{Ballesteros:1994fs}
A.~Ballesteros, F.~J.~Herranz, M.~A.~del Olmo and M.~Santander, 
{\it Four-dimensional quantum affine algebras and space-time q-symmetries}, 
J.\ Math.\ Phys.\ {\bf 35}, 4928 (1994).

\bibitem{Gutierrez:2021cs}
I.~Gutierrez-Sagredo and F.~J.~Herranz, 
{\it Cayley-Klein Lie Bialgebras: Noncommutative Spaces, Drinfel'd Doubles and Kinematical Applications}, 
Symmetry {\bf 13}, 1249 (2021) [arXiv:2106.03817 [hep-th]].

\bibitem{Ballesteros:2021is}
A.~Ballesteros, G.~Gubitosi and F.~Mercati,
{\it Interplay between Spacetime Curvature, Speed of Light and Quantum Deformations of Relativistic Symmetries}, 
Symmetry {\bf 13}, 2099 (2021) [arXiv:2110.04867 [gr-qc]].


\bibitem{Chari:1994as}
V.~Chari and A.~N.~Pressley, 
``A guide to quantum groups,'' 
Cambridge Univ.\ Press 1994.

\bibitem{Majid:1995fy}
S.~Majid, 
``Foundations of Quantum Group Theory,'' 
Cambridge Univ.\ Press 1995.

\bibitem{Papageorgiou:2009as}
G.~Papageorgiou and B.~J.~Schroers, 
{\it A Chern-Simons approach to Galilean quantum gravity in 2+1 dimensions}, 
J.\ High Energy Phys.\ {\bf 11}, 009 (2009) [arXiv:0907.2880 [hep-th]].

\bibitem{Papageorgiou:2010ga}
G.~Papageorgiou and B.~J.~Schroers, 
{\it Galilean quantum gravity with cosmological constant and the extended $q$-Heisenberg algebra}, 
J.\ High Energy Phys.\ {\bf 11}, 020 (2010) [arXiv:1008.0279 [hep-th]].

\bibitem{Hartong:2016ny}
J.~Hartong, Y.~Lei and N.~A.~Obers, 
{\it Nonrelativistic Chern-Simons theories and three-dimensional Ho\v{r}ava-Lifshitz gravity}, 
Phys.\ Rev.\ D {\bf 94}, 065027 (2016) [arXiv:1604.08054 [hep-th]].

\bibitem{Fontanella:2022cs}
A.~Fontanella and S.~J.~van Tongeren, 
{\it Coset space actions for nonrelativistic strings}, \linebreak
J.\ High Energy Phys.\ {\bf 10}, 80 (2022) [arXiv:2203.07386 [hep-th]].

\bibitem{Donnay:2023by}
L.~Donnay, A.~Fiorucci, Y.~Herfray and R.~Ruzziconi, 
{\it Bridging Carrollian and celestial holography}, 
Phys.\ Rev.\ D {\bf 107}, 126027 (2023) [arXiv:2212.12553 [hep-th]].

\end{thebibliography}
\end{document}